\documentclass[11pt,a4paper]{article}  	
\usepackage{geometry}                		

\usepackage{amssymb,amsmath}
\usepackage{float}
\usepackage{tabularx}
\usepackage{graphicx}
\usepackage{t1enc}
\usepackage[icelandic,english]{babel}
\usepackage{graphicx}
\usepackage{amsmath}
\usepackage{amssymb}
\usepackage{multirow}
\usepackage{enumitem}
\usepackage{t1enc}
\usepackage{amssymb}
\usepackage{color}
\usepackage{latexsym}
\usepackage{epic}
\usepackage{epsfig}
\usepackage{eepic}
\usepackage{amsmath}
\usepackage{enumerate}
\usepackage{graphicx}
\usepackage{multicol}	
\usepackage{natbib}
\usepackage{tabularx}
\usepackage{graphicx}
\usepackage{caption}
\usepackage{subcaption}
\numberwithin{equation}{section}

\newcommand{\Matrix}[1]{\begin{pmatrix} #1 \end{pmatrix}}

\newcommand{\fat}[1]{\boldsymbol{#1}}

\newcommand{\inv}{^{-1}}
\newcommand{\ndist}[2]{\mathcal{N}\left(#1,#2\right)} 
\newcommand{\trp}{^\mathsf{T}}

\newcommand{\ei}{\end{itemize}}

\usepackage{subcaption}
\usepackage{algorithm}
\usepackage{algorithmic}

\author{ Óli Páll Geirsson and  Birgir Hrafnkelsson \\
      Department of Mathematics\\
      Faculty of Physical Sciences\\
    School of Engineering and Natural Sciences\\
    University of Iceland, \\Iceland
    \ \\
      \ \\
    and
    \ \\
     \ \\
    Daniel Simpson \\
     Department of Mathematical Sciences\\
    Norwegian University of Science and Technology\\
    Norway\\}
\title{Computationally efficient spatial modeling of annual maximum 24 hour precipitation. An application to data from Iceland.}
\begin{document}
\maketitle

\begin{abstract}
We propose a computationally efficient statistical method to obtain distributional properties of annual maximum 24 hour precipitation on a 1 km by 1 km regular grid over Iceland. A latent Gaussian model is built which takes into account observations, spatial variations and outputs from a local meteorological model.  A covariate based on the meteorological model is constructed at each observational site and each grid point in order to assimilate available scientific knowledge about precipitation into the statistical model. The model is applied to two data sets on extreme precipitation, one uncorrected data set and one data set that is corrected for phase and wind.  The observations are assumed to follow the generalized extreme value distribution. At the latent level, we implement SPDE spatial models for both the location and scale parameters of the likelihood. An efficient MCMC sampler which exploits the model structure is constructed, which yields fast continuous spatial predictions for spatially varying model parameters and quantiles. 
\end{abstract}

\noindent \textbf{Address for correspondence:} O.P. Geirsson, Department of Mathematics, School of Engineering and Natural Sciences, University of Iceland, Dunhagi 5, 107 Reykjavík, Iceland. E-mail: \texttt{olipalli@gmail.com}



\section{Introduction}
Extreme rainfall in Reykjavík on the 16th of August 1991 resulted in overloaded local drainage systems causing severe damage to industrial buildings, houses and apartments. Even though extreme precipitation events are rare, understanding their frequency and intensity is important for public safety and various types of long term agricultural, industrial and urban planning. 

Meteorological models, of various complexities, have been proposed to model precipitation. For mountainous regions, such as much of Iceland, meteorological models that take into account the effects of topography as well as atmospheric information on orographic precipitation are needed. To that extent, \cite{smith2004linear} proposed a quasi-analytic linear orographic method to simulate precipitation over a spatial domain which \cite{crochet2007estimating} adopted to precipitation in Iceland. The resulting model simulates daily precipitation on a 1 km by 1 km regular grid across Iceland over the years 1958-2002.

Although meteorological models are well suited to give insight into the spatial mean behavior of precipitation, they tend to deviate significantly from observations when predicting extreme precipitation. The first goal of this paper is to propose a statistical method that takes into account observations, spatial variation of the extremes and outputs from regional meteorological models to increase predictive power of extreme events. To that extent, a covariate based on the meteorological model of \cite{crochet2007estimating} is constructed in order to assimilate much of the scientific knowledge about precipitation into a statistical model.

In recent years, various statistical models have been proposed for modeling extreme precipitation, where models based on the generalized extreme value distribution are standard in the literature,  see for example \cite{sang2009hierarchical}. Furthermore, the Bayesian approach is well suited to quantify uncertainty of the underlying physical processes as further argued in \citep{tebaldi2009joint}. In particular, the spatial variation can be modeled through the likelihood parameters as presented in e.g. \citep{davison2012statistical, hrafnkelsson2012spatial}. Alternative modeling approaches have been explored, such as the peek over threshold methods  with Generalized Pareto Distribution \citep{cooley2007bayesian}. 

Latent Gaussian models (LGMs) \citep{rue2005gaussian}, which form a flexible and practical subclass of Bayesian hierarchical models, play a dominant role in the vast literature on spatial statistics \citep{delfiner2009geostatistics, diggle1998model, guttorp2006studies}. In the LGM framework, Gaussian fields (GFs) appear at the latent level of the hierarchical model. However, posterior inference for GFs becomes increasingly computationally demanding as data sets get larger. Gaussian fields can be approximated by Gaussian Markov random fields (GMRFs), which increases the speed of computation significantly \citep{rue2001fast}. Although GMRFs are computationally efficient they become difficult to parameterize in a spatial setting. In recent work by \cite{lindgren2011explicit},  numerical approximations to GFs with Matérn covariance structure are presented. The method constructs an approximation  of a stochastic partial differential equation (SPDE)  on a triangulated mesh. The approximate solutions can then be used to construct a GMRF representation of the desired GF on the mesh. This allows for continuous spatial predictions by choosing appropriate basis functions for the approximation.

The second goal of the paper is to present a computationally efficient Bayesian hierarchical spatial model in order to obtain distributional properties of extreme precipitation on a high resolution grid. To that extent, an LGM is presented where the observations are assumed to follow the generalized extreme value distribution. The spatial variations are modeled through the location and scale parameters of the likelihood with SPDE spatial models. An MCMC split sampler \citep{opgsplit} is applied to the model structure, yielding an efficient inference scheme.  Furthermore, the model structure can also be used to make spatial predictions for the model parameters and quantiles of extreme precipitation on the high resolution grid.

The paper is organized as follows. The data and the meteorological model are presented in Section 2, where a method to construct covariates from outputs of the meteorological model is also outlined. Detailed description of the model structure, the inference method and spatial prediction is given in Section 3. Posterior results are presented and discussed in Section 4. The paper concludes with a discussion in Section 5.


\section{The data}

\subsection{Observations}

\begin{figure}[htp]
\centering
	\begin{subfigure}[b]{0.475 \textwidth}
    	\includegraphics[width=\textwidth, trim =  0mm 10mm 10mm 10mm, clip]{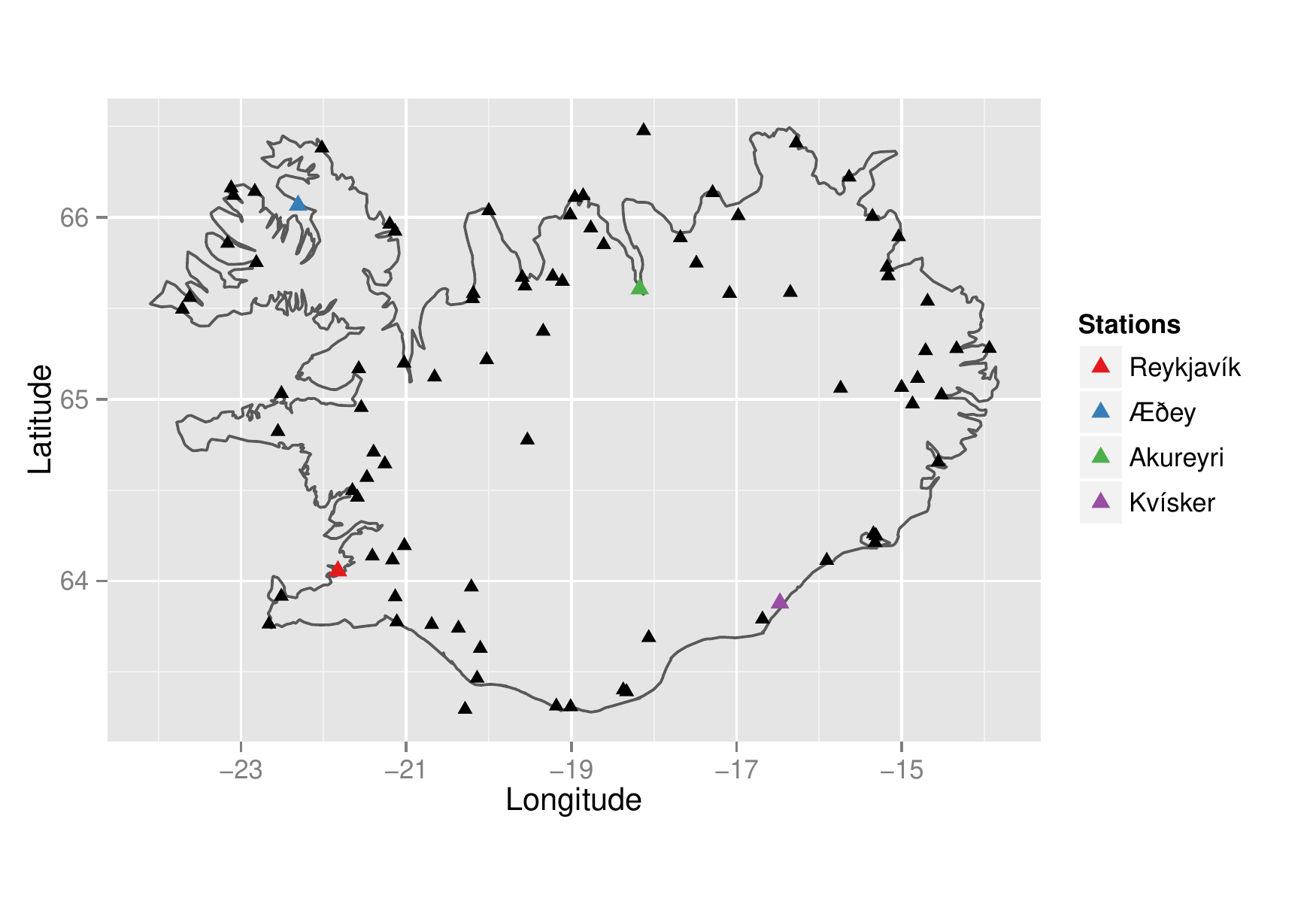}
  	\end{subfigure}
  	\begin{subfigure}[b]{0.475 \textwidth}
    	\includegraphics[width=\textwidth, trim =  0mm 10mm 10mm 10mm, clip]{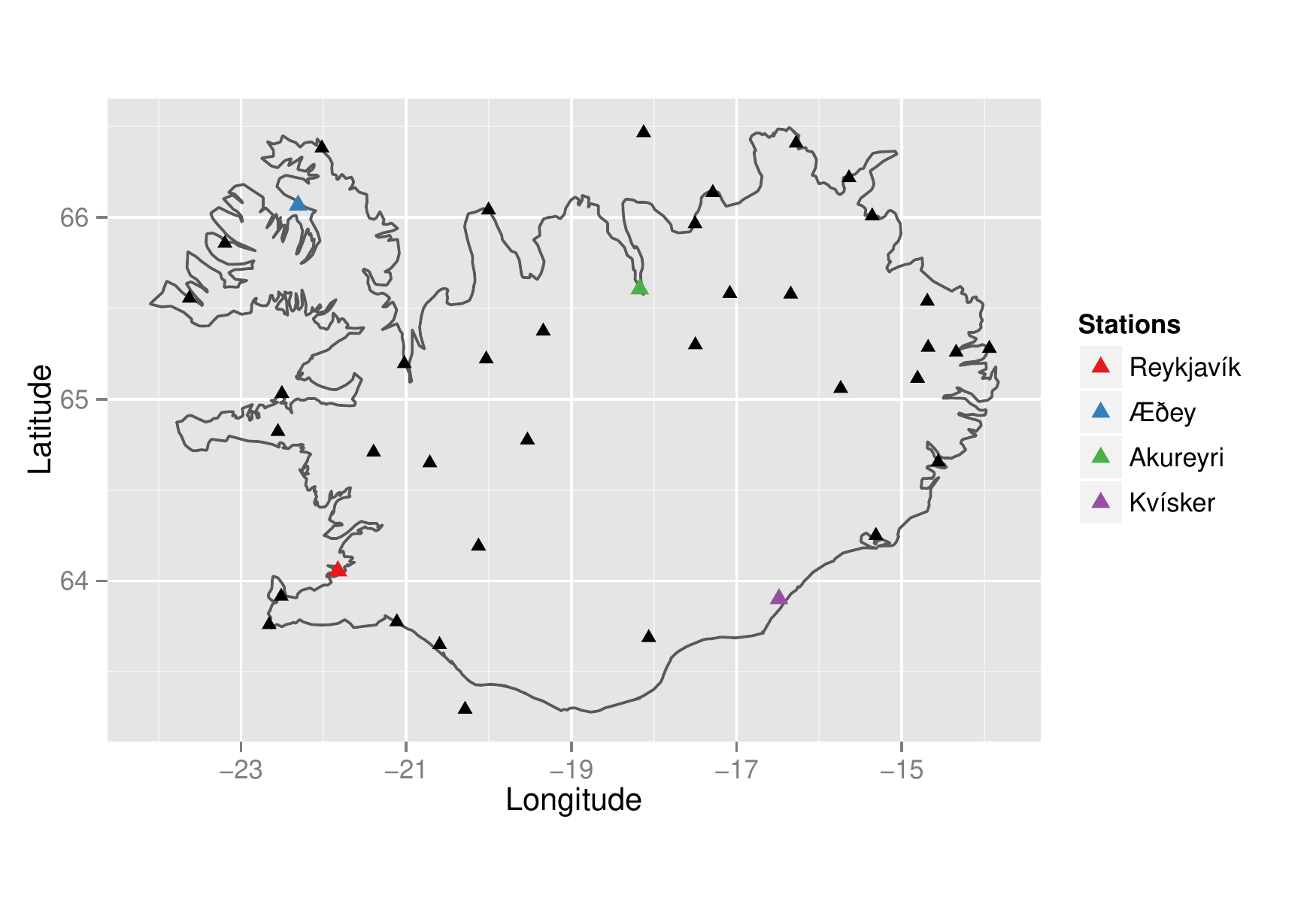}
  	\end{subfigure}
\caption{Observational sites. Left and right panel show the locations in the uncorrected and corrected data sets  respectively}
   \label{FIG:obs}
\end{figure}

The observed data on precipitation were provided by the Icelandic Meteorological Office (IMO). Two types of data sets were explored in this paper, referred to as uncorrected and corrected data sets. The uncorrected data set contains raw observations of 24 hour annual maximum precipitation from 86 sites in Iceland over the years 1958 to 2006, as seen in \citep{crochet2007estimating}. The latter data set is based on observations that have been have been corrected according to the dynamic correction model proposed by \citep{forland1996manual}. The correction accounts for trace; wetting and evaporation losses; and for the catch deficiencies due to aerodynamic effects. \cite{crochet2007estimating} made corrections for 40 observational sites across Iceland over the years 1958 to 2006 on a daily basis, according to the phase (liquid, mixed, solid) of hydrometeors and the average daily wind speed and temperature recorded at each site. The observational sites for both data sets can be seen on Figure \ref{FIG:obs}, where the uncorrected and corrected data sets are on the left and right panels,  respectively.

\begin{figure}[htp]
   \centering
   \centerline{\includegraphics[width=1.15 \textwidth]{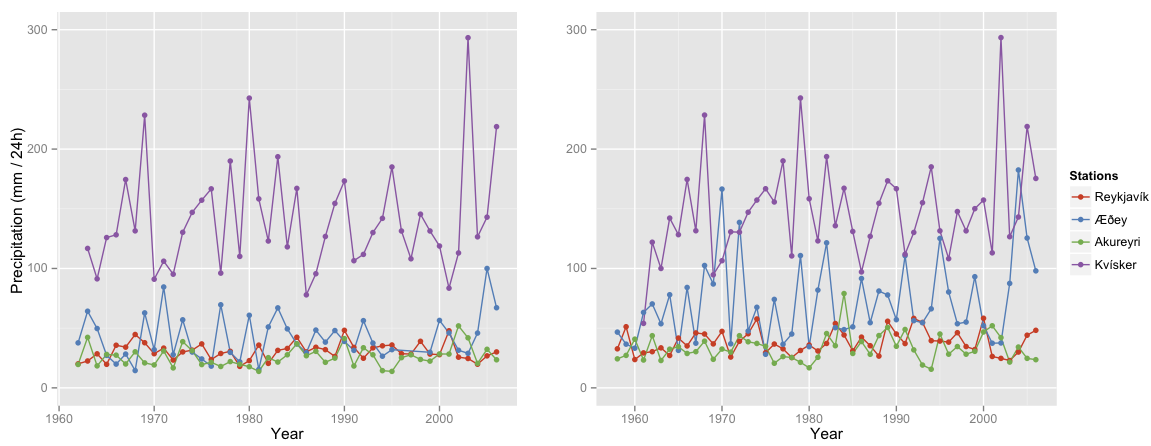}} 
   \caption{Times series. Left and right panel show the time series  in the uncorrected and corrected data sets  respectively}
   \label{FIG:timseS}
\end{figure}

Time series from four observational sites, Reykjavík, Æðey, Akureyri and Kvísker, are shown on Figure \ref{FIG:timseS} for the uncorrected and corrected data sets. Reykjavík and Akureyri were chosen because they are the most populated areas in Iceland. Æðey was chosen to demonstrate the difference between the uncorrected and corrected data sets. Finally, Kvísker was chosen as it has the highest observed precipitation, and is by nature difficult to infer as it exhibits higher extremal behaviour than its neighbours.

\subsection{The meteorological model}

The meteorological model that is used in this paper to improve spatial predictive power, is based on a linear orographic precipitation model proposed by \cite{smith2004linear}. \cite{crochet2007estimating} have adapted the method to precipitation in Iceland. The model is driven by coarse resolution precipitation, wind and temperature data obtained from re-analyses (1958-2001) (ERA-40) \citep{uppala2005era} and analyses (2002-2004) made by the European Center for Medium range Weather Forecast. The model takes into account the topography of the spatial domain, airflow dynamics, condensed water advection and downslope evaporation. This means that outputs from the model contains information about the underlying physical processes of precipitation. Note that, three free parameters of the meteorological models were calibrated using five years worth of daily precipitation data at forty observational sites across Iceland and precipitation information from three ice caps.
 
 The resulting model simulates daily precipitation on a 1 km by 1 km regular grid across Iceland for the years 1958-2002.  The study domain is on a $521 \text{ km} \times 361 $ km regular grid. Thus, the simulated data set is (roughly) of the magnitude $3\cdot 10^{9}$ data points, as the spatial dimension is $521 \times 361$ and the temporal dimension is $365 \times 44$.

\subsection{Covariates}

Reasonable covariates for extreme precipitation should assimilate available spatial information such as the topography of the domain and the underlying physical processes of precipitation in order to increase the predictive power of the spatial predictions. By constructing covariates based on the meteorological model, the information about the above factors can be assimilated. Furthermore,  \cite{benestad2012spatially}  suggested that observed mean values of precipitation have high predictive power for extreme precipitation. Extending that argument, we propose that the calculated sample mean values from the meteorological model will have similar predictive power to the observed means.  Moreover, the simulated data contain knowledge of the underlying physical processes. This leads to a high-quality, data consistent predicted mean over the entire spatial domain. This is a novel extension of the concept of \cite{benestad2012spatially} to spatial prediction of extremes. The means based on the meteorological model are referred to hereafter as simulated means. 

The simulated means can be calculated at every grid point, which yields a {521 km $\times $ 361 km} regular grid of covariates. However, observational sites are not necessarily at the regular grid points. In order to construct the covariates at the observational sites we implemented the following spatial smoother.  

\begin{figure}[htp]
   \centering
   \centerline{\includegraphics[width=0.65 \textwidth]{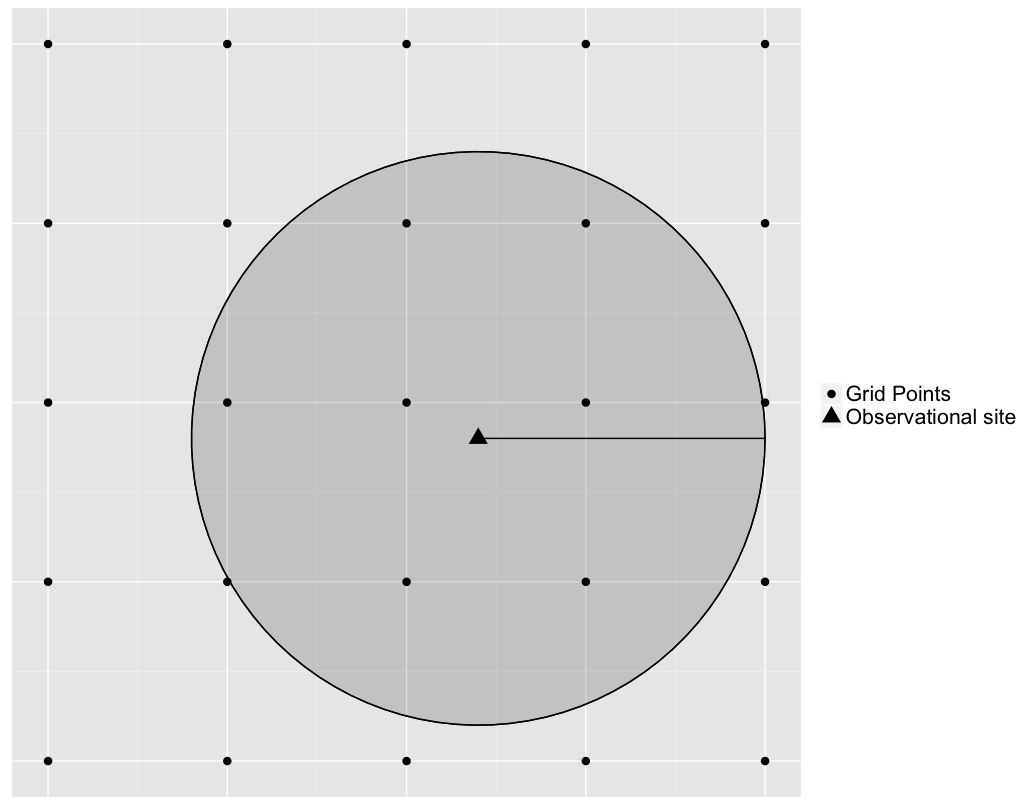}} 
   \caption{Illustrative figure, how the index set $D(r,i)$ is created for some radius $r$ and observational site $i$. All the grid points within the circle then belong to $D(r,i)$}
   \label{FIG:circle}
\end{figure}

Let $\mathcal{G}$ denote the set of every regular grid point and let $\mathcal{S}$ denote the set of the  observational sites.  Furthermore, let  $\bar x^*_j $ denote the simulated means at every grid point $j \in \mathcal{G}$ and $ \bar x_i$ denote observed means at each observational site $i \in \mathcal{S}$. For a given tuning distance $r>0$, find every grid point $j \in \mathcal{G}$ that is within distance $r$ of point $i \in \mathcal{S} $ and denote that index set with $D(r,i)$. Illustrative figure can been seen in Figure \ref{FIG:circle}.

The goal is to construct a covariate at site $i$ that uses information from the grid points in $D(r,i)$. Assign the following decay function
\[
	 w_{ij} = 1 - F_{\alpha}\left(\frac{d_{ij}}{r}\right) \quad \text{ for all $j \in D(r,i)$}
\]
where $F_{\alpha}$ denotes the cumulative density function of the beta distribution with parameters $\alpha$ and $\alpha$, and  $d_{ij}$ denotes the Euclidean distance from point $i$ to point $j$. The two parameters of the beta distribution are chosen to be the same in order to avoid interactions between the tuning distance $r$ and the shape of the decay function. The $w_{ij}$s describe the relative weights of the neighboring grid points in $D(r,i)$. The covariate at observation point $i$ can then be constructed as the weighted mean of the surrounding simulated means, that is
\begin{equation}
\label{weight}
	\bar {x }^*_i(r,\alpha) = \frac {\underset{j \in D(r,i) }{\sum} w_{ij}{\bar x}^*_j} {\underset{j \in D(r,i) }{\sum} w_{ij}} \quad \text{for every $i$ in $\mathcal S$.}
\end{equation}
Following the \citep{benestad2012spatially} argument, the parameters in the spatial smoother are tuned such that the simulated means at the observational sites are close to the observed means. We chose to measure the distance between the simulated means, which are calculated with equation (\ref{weight}), and the observed means by using mean square distance. Then the optimal $r$ and $\alpha$ can be found in the mean square sense by calculating the following
\[ 
	(r_0, \alpha_0) = \arg\min_{r,\alpha} \sum_{i\in \mathcal{S}} || \bar {x }^*_i(r,\alpha) - \bar x_i ||^2.
\]
The covariates that will be used in this paper are then $\bar {x }^*_i = \bar {x }^*_i(r_0,\alpha_0)$ for all observational sites $i \in \mathcal S$.

\section{The model and inference}
\cite{guttorp2006studies} proposed that using Gaussian fields with a Matérn covariance structure, called Matérn fields, is a flexible and interpretable way of modeling underlying physical processes of natural phenomena in a spatial setting. However, the resulting covariance matrices are dense and thus become computationally demanding in posterior inference and spatial predictions as data sets get larger.

Approximating Gaussian fields as Gaussian Markov random fields (GMRF) \citep{knorr2002block} provides an efficient framework for computationally efficient Gaussian models. GMRFs are parameterized with precision matrices, which are defined as inverses of covariance matrices. Although GMRFs have very good computational properties, there was no standard way to parametrize the precision matrix of a GMRF to achieve a predefined spatial covariance structure between two sites until \cite{lindgren2011explicit}  addressed the issue.  They proposed a method which computes a numerical approximation to the Matérn field on a triangulated mesh over the spatial domain, based a stochastic partial differential equation (SPDE) representation.  The approximate solution $x(\fat s)$, for each point $\fat s$ in the spatial domain,  is on the form
\begin{align}
\label{appr}
	x(\fat s ) = \sum_{k=1}^n \psi_k(\fat s_k)w_k
\end{align}
where $n$ is the number of vertices in the mesh, $\psi_k$ are piecewise linear basis functions and $w_k$ are Gaussian weights. The approximate solutions can be used to construct a GMRF representation of the Matérn field at every point within the triangulated mesh.

In order to model the behavior of the underlying physical processes of extreme precipitation, the location and scale parameters in the likelihood are allowed to vary in space. To that extent, SPDE spatial models are used at the latent level of the proposed hierarchical model to describe continuously the spatial variation of these two latent parameters. Posterior inference and spatial predictions based on this approach are presented.

\subsection {Model structure}

The data are modeled with an LGM assuming the generalized extreme value distribution for the observations. That is, let $y_{it}$ denote the annual maximum 24 hour precipitation at station $i$ at year $t$, with a cumulative density function of the form
\begin{equation*}
	F(y_{it}) =  \exp\left\{ - \left(1+ \xi\left(\frac{y_{it}-\mu_i}{\sigma_i}\right)\right)^{-1/\xi} \right\},\quad i = 1,\ldots, J,\quad t=1,\ldots,T
\end{equation*}
if  $1+\xi(x-\mu)/\sigma>0$, $F(y_{it}) = 0$ otherwise. The parameters $\mu_{i}$, $\sigma_i $ and $\xi$ are location, scale and shape parameters; $J$ is the number of observational sites and $T$ is the number of years. These distributional assumptions are reasonable as the generalized extreme value distribution belongs to the family of extremal distributions and have desired asymptotic properties \citep{coles2001introduction}. The observations are assumed to be  independent conditioned on latent parameters. The conditional independence assumption implies that realizations of the spatial maxima will be everywhere discontinuous. This is somewhat unrealistic and may affect the width of the posterior intervals.  On the other hand,  most sites are far enough apart that they can be considered conditionally independent. Furthermore, the main purpose is to give spatial predictions of marginal quantiles using the SPDE approach, but not to simulate spatial realizations of the extreme precipitation for the next year or unobserved sites for an observed year.  Hence, models with data-level dependence are not considered in this paper. 

At the latent level of the model, we implement the following spatial model structure for the location parameter  $\fat \mu = (\mu_1,\ldots, \mu_J)$, 
\begin{equation}
\label{modelstr}
	\fat \mu = \fat X_\mu \fat{\beta_\mu} + \fat A_{\mathcal{S}} \fat u_\mu + \fat v_\mu,
\end{equation}
where $\fat X_\mu$ is a design matrix consisting of a vector of ones and the covariates $\bar {x }^*_i$ that are based on the meteorological model; $\fat{\beta_\mu}$ are the corresponding weights; $\fat u_\mu $ is a spatial effect on a triangulated mesh; $\fat A_{\mathcal{S}}$ is a projection matrix; the matrix product $\fat A_{\mathcal{S}}\fat u_\mu $ then denotes the spatial effect at the observational sites, which captures the spatial variation in the data that is unexplained by the covariate and $\fat v_\mu$ is an unstructured random effect.

\begin{figure}[h!] 
   \centering
   \includegraphics[width=0.95 \linewidth, trim = 20mm 30mm 20mm 30mm, clip]{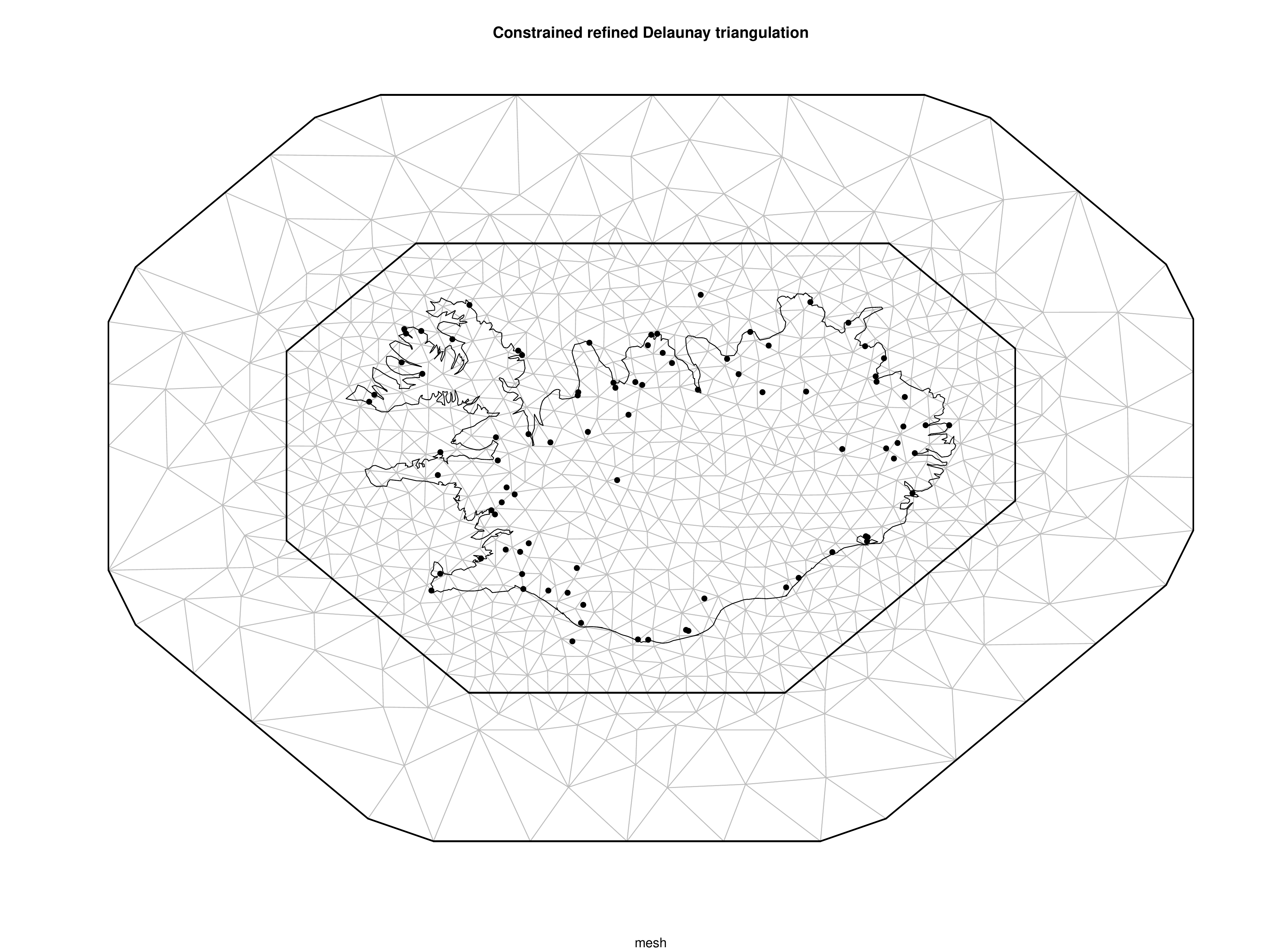}
   \caption{Triangulated mesh over Iceland}
   \label{FIG:triangles}
\end{figure}

The spatial effect $\fat u_\mu$ is constructed using the SPDE approach. In order to obtain the SPDE structure for $\fat u_\mu$, the domain of interest is subdivided into a mesh of non-intersecting triangles, see Figure \ref{FIG:triangles}. The  GMRF representation is then constructed for the weights of the basis functions in \eqref{appr}  with a precision matrix $\fat Q_{u \mu}(\fat \psi_{u \mu })$ constructed with the SPDE approach of \cite{lindgren2011explicit}. The precision matrix is based on the geometry of the mesh and is, by construction, a sparse matrix. In this paper, we chose the smoothness parameter to be $\nu = 1$, which corresponds to an almost once differentiable Matérn field. The precision matrix $\fat Q_{ u \mu}(\fat \psi_{u \mu})$ has two parameters, $\fat \psi_{u \mu}=(\kappa_{u \mu}, \omega_{u \mu})$, which enter the model as hyperparameters. The hyperparameter  $\kappa_{u \mu}$ is inversely proportional to the range of the approximate  Matérn field and the hyperparameter $\omega_{u \mu}$ is related to the marginal variance.

The locations of $\fat u_\mu$ in space are on the vertices of the mesh, which are not necessarily at the observational sites. However, since the approximate representation (\ref{appr}) is assumed to have piecewise linear basis functions, an approximate solution can also be found within each triangle as a convex linear combination of the approximate solutions at the surrounding vertices. The spatial effect $\fat u_\mu$ can be projected linearly onto every point within every triangle of the mesh. The matrix $\fat A_{\mathcal S}$ denotes the linear projection from the mesh vertices onto the observational sites. Note that the number of non-zero entries in each line of $\fat A_{\mathcal S}$  is three and the sum of every line is one. 

Analogous spatial structure is also implemented for the scale parameter $\sigma_i$ on a logarithmic scale. That is, let $\tau_i = \log  \sigma_i$ and then model $\fat \tau = (\tau_1, \ldots, \tau_J)$ as
\[
	\fat \tau = \fat X_\tau \fat \beta_\tau + \fat A_{\mathcal{S}} \fat u_\tau + \fat v_\tau.
\]
However, the design matrix $\fat X_\tau$ consists of a vector of ones and a covariate based on the meteorological model on a logarithmic scale, that is $\log \bar {x }^*_i$.

\subsection{Prior selection}

The following prior distributions were assigned for the latent parameters,
\begin{align*}
\label{prior}
	\fat \beta_\mu &\sim \mathcal{N}(\fat 0, \kappa_{\beta \mu}^{-1} \fat I),  &  \fat \beta_\tau &\sim \mathcal{N}(\fat 0, \kappa_{\beta \tau}^{-1} \fat I),  \nonumber \\
	\fat u_\mu &\sim \mathcal{N}(\fat 0 , \fat Q^{-1}_{ u \mu}(\fat \psi_{u \mu})), & \fat u_\tau &\sim \mathcal{N}(\fat 0 , \fat Q^{-1}_{u \tau}(\fat \psi_{u \tau})),\\
	\fat v_\mu &\sim \mathcal{N}(\fat 0, \kappa_{v \mu}^{-1} \fat {I}), & \fat v_\tau &\sim \mathcal{N}(\fat 0, \kappa_{v \tau }^{-1} \fat {I}). \nonumber
\end{align*}
The parameters  $\fat \beta_\mu$ and $\fat \beta_\tau$ are assumed \emph{a priori} to have a low precision on their native scales in order to let the data play the dominate role in their inference. Thus, the parameter values $\kappa_{\beta \mu} = 0.0025$ and $\kappa_{\beta \tau}= 0.25$ were chosen for the prior distributions. 

Lognormal prior distributions with fixed parameters were assigned to the hyperparmeters of the spatial fields $\fat u_\mu $ and $\fat u_\tau$, that is
\begin{align*}
	\kappa_{u \mu} &\sim \mathcal {LN}\left(\mu_{\kappa u \mu},  \sigma^2_{\mu\kappa} \right), \quad \quad \kappa_{u \tau} \sim \mathcal {LN}\left(\mu_{\tau\kappa},  \sigma^2_{\tau\kappa} \right) \\
	\omega_{u \mu} &\sim \mathcal{LN}\left(\mu_{\mu \omega} ,  \sigma^2_{\mu \omega} \right), \quad \quad \omega_{u \tau} \sim \mathcal{LN}\left(\mu_{\tau\omega},  \sigma^2_{\tau\omega} \right)
\end{align*}
where $\mathcal {LN}$ denotes the lognormal distribution. The fixed parameters in the priors for $\kappa_{u \mu}$ and $\kappa_{u \tau}$ were chosen in a weakly informative manner to ensure that the range of the spatial effects is sensible relative to the size of domain of interest. The relationship between the parameters $\kappa_{u \mu}$ and $\kappa_{u \tau}$ and the spatial range of the approximated Matérn field used in this paper is $\rho = \sqrt{8\nu}/\kappa$, which corresponds to correlation near 0.1 at a distance $\rho$. Consequently, the prior distributions for $\kappa_{u \mu}$ and $\kappa_{u \tau}$ were chosen such that the corresponding $\rho$'s range up to half of the length of the domain, which in this paper is approximately 150 km. To that extent, the values $\mu_{\mu\kappa}=-2.7$,  $\sigma_{\mu\kappa}=0.45$, $\mu_{\tau\kappa}=-2.5$,  $\sigma_{\tau\kappa}=0.45$ were chosen.

Site specific maximum likelihood estimates for $\fat \mu$ and $\fat \tau$ in an exploratory data analysis and the relation to their corresponding covariates $\fat X_\mu$ and $\fat X_\tau$ suggest that the marginal standard deviations for $\fat u_\mu$ and $\fat u_\tau$  should exceed $50$ and $5$, respectively, with low probability on their native scales. The relationship between the parameters $\omega_{u \mu}$ and $\omega_{u \tau} $ and the marginal standard deviation of the Matérn field is approximately $1/(\sqrt{4\pi}\kappa \omega)$. Which led to the parameter values $\mu_{\mu\omega}=-1.1$,  $\sigma_{\mu\omega}=0.1$, $\mu_{\tau\omega}=1.8$,  $\sigma_{\tau\omega}=0.1$.

The unstructured random effects $\fat v_\mu$ and $\fat v_\tau$ capture the variation in the data which is unexplained by the covariates and the spatial models. The following prior distributions for the precision parameters of $\fat v_\mu$ and $\fat v_\tau$ were assigned
\begin{align*}
	\kappa_{v \mu} &\sim \mathcal {LN}\left(\mu_{\mu v},  \sigma^2_{\mu v} \right), \quad \kappa_{v \tau} \sim \mathcal {LN}\left(\mu_{\tau v},  \sigma^2_{\tau v} \right).
\end{align*}
The standard deviations of $\fat v_\mu$ and $\fat v_\tau$ are \emph{a priori} believed to be mainly between $2$ and $20$, and $0.1$ and $1.5$, respectively. Hence, the values  $\mu_{\mu v}=-4.5$,  $\sigma_{\mu v}=0.45$, $\mu_{\tau v}=1.5$, and $\sigma_{\tau v}=0.1$ were chosen.

Plots of the prior distributions for all the hyperparameters $\kappa_{u \mu}$, $\kappa_{u \tau}$, $\omega_{u \mu}$, $\omega_{u \tau}$, $\kappa_{v \mu}$ and $\kappa_{v \tau}$ are shown on transformed scales for interpretability. That is, the range of the two spatial fields are shown in Figures \ref{PP_RangeEta} and \ref{PP_RangeTau},  which are functions of $\kappa_{u \mu}$ and $\kappa_{u \tau}$; the marginal standard deviation of the spatial fields are shown in Figures \ref{PostPrior_SdEta} and \ref{PostPrior_SdTau}, which are functions of $\omega_{u \mu}$ and $\omega_{u \tau}$; and standard deviation of the unstructured random effects are shown in Figures \ref{PostPriorVloc} and \ref{PostPrior_VTau} , which are functions of $\kappa_{v_\mu}$ and $\kappa_{v_\tau}$.

Due to lack of data and for illustration purposes, the shape parameter $\xi$ was assumed constant as opposed to having it varying over the spatial domain.  Thus, the shape parameters was assigned the prior distribution
\[
	\xi \sim \mathcal N(0, \kappa_\xi^{-1})
\]
where $\kappa_\xi=2$. Note that $\xi$ is inferred in the model as a latent parameter as seen in \cite{hrafnkelsson2012spatial}.

\subsection{Posterior inference}
\label{SplitSampler}

Due to the proposed model structure, in particular the spatial model on the scale parameters, MCMC methods were necessary to make posterior inference as opposed to approximation methods such as INLA \citep{rue2009approximate}. However, standard MCMC methods like single site updating converged slowly and mixed poorly as many model parameters were heavily correlated in the posterior. To address this issue, posterior inference was done by using the MCMC split sampler \cite{opgsplit}. 

In order to implement the MCMC split sampler, the model parameters are set up as follows. First,  define the data-rich part $\fat \eta=(\fat\mu, \fat\tau, \xi)$ and the data-poor part $\fat \nu=(\fat\beta_\mu, \fat u_\mu,\fat \beta_\tau , \fat u_\tau)$. The latent field of the proposed LGM can then be written as
\[
	\fat x =(\fat \eta, \fat \nu)= (\fat \mu, \fat \tau, \xi, \fat \beta_{\mu}, \fat u_{\mu}, \fat \beta_{\tau}, \fat u_{\tau})
\]
with hyperparameters, $\fat \theta=(\kappa_{u \mu}, \omega_{u \mu},\kappa_{u \tau}, \omega_{u \tau}, \kappa_{v \mu}, \kappa_{v \tau})$. Then define

\[
	\fat Z = \begin{pmatrix} 
		\fat X_\mu & \fat A_{\mathcal{S}}  & \cdot & \cdot  \\
		\cdot & \cdot & \fat X_\tau & \fat A_{\mathcal{S}}   \\
		\cdot & \cdot & \cdot & \cdot \\
	\end{pmatrix}
	\text{, }
	\quad
	\fat Q_\epsilon = \begin{pmatrix} 
		\kappa_{v \mu} \fat I  & \cdot & \cdot \\
		\cdot & \kappa_{v \tau} \fat I  & \cdot \\
		\cdot & \cdot & \kappa_\xi \\
	\end{pmatrix}
	\text{, }
	\quad
	\fat Q_\nu = \begin{pmatrix} 
		\kappa_{\beta \mu} \fat I  & \cdot & \cdot & \cdot \\
		\cdot  & \fat Q_{ u \mu} & \cdot & \cdot \\
		\cdot  & \cdot & \kappa_{\beta \tau} \fat I  & \cdot \\
		\cdot  & \cdot & \cdot  &  \fat Q_{ u \tau}\\
	\end{pmatrix}.
\]
where we use $\fat Q_{ u \mu} $ instead of $\fat Q_{ u \mu}(\fat \psi_{u \mu})$ for notational conveniance and the dotted entries denote zero elements. The joint prior distribution of the latent field conditioned on the hyperparmeters, i.e. $\fat x | \fat \theta $, is then a mean zero Gaussian with the following precision matrix
\begin{align*}
	\fat Q = \Matrix{\fat{Q}_\epsilon & -\fat{Q}_\epsilon\fat Z \\ -\fat Z\trp\fat{Q}_\epsilon & \fat{Q}_\nu+\fat Z\trp\fat{Q}_\epsilon\fat Z}.
\end{align*}
The corresponding posterior density is,
\begin{align*}
	\pi (\fat x, \fat \theta | \fat y ) & = \pi (\fat \eta,\fat \nu, \fat \theta | \fat y ) \propto \pi (\fat y | \fat \eta) \pi (\fat \eta, \fat \nu | \fat \theta )\pi(\fat \theta)
\end{align*}
A sample from the posterior density $\pi(\fat \eta, \fat \nu, \fat \theta | \fat y)$ is then obtained in with a two block Gibbs sampler type, and is briefly outlined below. The details can be seen in \cite{opgsplit}.

\subsubsection{ Data-rich part}
The structure of the data-rich part is as follows. Let $\fat \eta_A = (\fat \mu, \fat \tau)$. In order to exploit the conditionally independence structure of the model, we sample from  $\pi(\fat \eta_A | \fat y, \xi,\fat \nu, \fat \theta)$ and  $\pi(\xi | \fat y, \fat \eta_A,\fat \nu, \fat \theta)$  in two blocks.  To sample from $\pi(\fat \eta_A | \fat y, \xi,\fat \nu, \fat \theta)$, start by defning
\[
	\fat Z_A = \begin{pmatrix} 
		\fat X_\mu & \fat A_{\mathcal{S}}  & \cdot & \cdot  \\
		\cdot & \cdot & \fat X_\tau & \fat A_{\mathcal{S}} \\
	\end{pmatrix}
	\quad
	\text{, }
	\quad
	\fat Q_A = \begin{pmatrix} 
			\kappa_{v \mu} \fat I & \cdot \\
		\cdot & 	\kappa_{v \tau } \fat I
	\end{pmatrix}
\]
and
\[
	\fat \eta_{A(i)}=(\mu_i, \tau_i) \quad\text{for}\quad i=1,\ldots, J.
\] 
The logarithm of the conditional posterior becomes
\begin{align*}
	\log \pi ( \fat \eta_A | \fat y, \xi,\fat \nu, \fat \theta) = -\frac{1}{2} \fat \eta_A\trp \fat Q_A \fat \eta_A +(\fat Q_{A} \fat Z\fat \nu)\trp \fat \eta_A + f_A( \fat \eta_A) + \text{const} 
\end{align*}
where
\[
	f_A(\fat \eta_A)  = \sum_{i=1}^J f_i(\fat \eta_{A(i)}) = \sum_{i=1}^J \sum _{i \in \mathcal{A}_i} \log f_{\text{gev}}(y_{it} | \mu_i, \exp{\tau_i}, \xi),
\]
where $f_{\text{gev}}$ denotes the density of the generalized extreme value distribution and the set $\mathcal{A}_i$ contains the indices of the years $t$ observed at site $i$.  The steps of the sampler are outlined in Algorithm 1.

\begin{algorithm}[htp]
\label{alg_eta}
\caption{Obtain the $(k+1)$-th sample from $\pi ( \fat \eta_A | \fat y, \xi,\fat \nu, \fat \theta) $}
\begin{algorithmic}[1]
\REQUIRE $\fat \eta_A^k$
\STATE Find the mode $\fat \eta_A^0 = \underset{\fat \eta_A} {\operatorname*{arg\,max}} ~ \log \pi ( \fat \eta_A | \fat y, \xi,\fat \nu, \fat \theta) $
\STATE Calculate $\fat H = \nabla^2 f_A(\fat  \eta_A^0)$  and  $\fat b = \nabla f_A(\fat \eta_A^0) - \fat H \fat \eta_A^0$
\STATE Sample $\fat \eta_A^* \sim \mathcal{N} \left( \fat \eta^0_A, \left( \fat Q_A -\fat H \right)\inv \right)$ 
\STATE Calculate  $\fat \rho (\fat \eta^k)$ and $\fat \rho (\fat \eta^*)$, where $ \fat \rho(\fat \eta) = \left(-\frac1 2 \fat \eta\trp \fat H   - \fat b\trp \right) \circ \fat \eta$ and $\circ$ denotes an entrywise product  
\FOR{$i=1,\ldots, J$}
\STATE Calculate $r_i =f_i( \fat \eta_{(i)}^*) + \fat \rho (\fat \eta^*)_{(i)}\trp \fat 1 - \left(f_i( \fat \eta_{(i)}^k) + \fat \rho (\fat \eta^k)_{(i)}\trp \fat 1 \right)$ 
\STATE Calculate $\alpha_i = \min\left\{ 1, \exp r_i \right\}$
\STATE Sample $u_i \sim \mathcal{U}(0,1)$
\IF{$\alpha_i > u_i$}
    \STATE $\fat \eta^{k+1}_{A(i)} = \fat \eta^{*}_{A(i)} $
  \ELSIF{$\alpha_i < u_i$}
  \STATE $\fat \eta^{k+1}_{A(i)} = \fat \eta^{k}_{A(i)} $
 \ENDIF
\ENDFOR
\ENSURE $\fat \eta^{k+1}$
\end{algorithmic}
\end{algorithm}

In order to sample from  $\pi(\xi | \fat y, \fat \eta_A,\fat \nu, \fat \theta)$, note that the logarithm of the corresponding conditional posterior is
\begin{align*}
	\log \pi( \xi | \fat y, \fat \eta_A,\fat \nu, \fat \theta) &= \log \pi(\xi) + f_B(\xi) 
\end{align*}
where
\[
	f_B(\xi)  = \sum_{i=1}^J \sum _{i \in \mathcal{A}_i} \log f_{\text{gev}}(y_{it} | \mu_i, \exp{\tau_i}, \xi),
\]
The steps of the sampler are outlined in Algorithm 2.
\begin{algorithm}[htp]
\label{alg_xi}
\caption{Obtain the $(k+1)$-th sample from $\pi ( \xi | \fat y, \fat \eta_A,\fat \nu, \fat \theta) $}
\begin{algorithmic}[1]
\REQUIRE $\xi^k$
\STATE Find the mode $\xi^0 = \underset{\xi} {\operatorname*{arg\,max}} ~ \log \pi ( \xi | \fat y, \fat \eta_A,\fat \nu, \fat \theta) $
\STATE Sample $\xi^* \sim \ndist{\xi^0}{ (\kappa_\xi - f''(\xi^0))\inv}$ 
\STATE Calculate  $\rho (\xi^k)$ and $\rho (\xi^*)$, where $\rho(\xi) = \left(- \frac{f''(\xi^0)}2 \xi - \Big(f'(\xi^0) - f''(\xi^0)\xi^0\Big)\right)\xi$
\STATE Calculate $r = f_B(\xi^*) + \rho(\xi^*) - \left( f_B(\xi^k) + \rho(\xi^k)  \right)$
\STATE Calculate $\alpha = \min\left\{ 1, \exp r \right\}$
\STATE Sample $u \sim \mathcal{U}(0,1)$
\IF{$\alpha > u$}
    \STATE $\xi^{k+1} = \xi^*$
  \ELSIF{$\alpha < u$}
  \STATE $\xi^{k+1}  = \xi^{k} $
 \ENDIF
\ENSURE $\xi^{k+1}$
\end{algorithmic}
\end{algorithm}

\subsubsection{ Data-poor part}
The proposal strategy suggested in \citep{knorr2002block} is used for each element of $\fat \theta$. That is, let $\theta_i^* = f\theta_i^k$ where the scaling factor $f$ has the density
\begin{equation}
\label{HavardProp}
	\pi(f) \propto 1 + 1/f  \quad \text{for $f \in [1/F, F]$}
\end{equation}
where $F>1$ is a tuning parameter. It can be shown that this a symmetric proposal density in the sense that
\[
	q(\theta_i^*|\theta_i^k) = q(\theta_i^k|\theta^*_i)
\]
A joint proposal strategy for $(\fat \nu, \fat \theta)$ is implemented as seen in \cite{opgsplit}. The steps of the sampler are outlined in Algorithm 3.

\begin{algorithm}[htp]
\label{alg_nu}
\caption{Obtain the $(k+1)$-th sample from $\pi (\fat \nu, \fat \theta  | \fat y, \fat \eta ) $}
\begin{algorithmic}[1]
\REQUIRE $(\fat \nu^k, \fat \theta^k)$
\STATE  Sample each element of $\fat \theta^*$ from the proposal density in (\ref{HavardProp})
\STATE Sample $\fat \nu ^*$ from
\begin{equation*}
\label{NonLikTheta}
	\fat \nu^* | \fat \eta,\fat \theta^{*} \sim \mathcal N  \left( \left(  \fat{Q}_\nu+\fat Z\trp\fat{Q}_\epsilon\fat Z \right)\inv \fat Z\trp \fat Q_\epsilon \fat\eta , \left( \fat{Q}_\nu+\fat Z\trp\fat{Q}_\epsilon\fat Z \right)\inv \right)
\end{equation*}
\STATE Calculate 
\begin{align*}\label{gmrfratio}
	r = \frac{\pi (\fat \theta ^*)}{\pi(\fat \theta ^k)}\cdot
	\frac{\pi( \fat \eta | \fat \nu^*, \fat \theta^*) \pi( \fat \nu^* | \fat \theta^*) }{\pi( \fat \nu^* |\fat \eta, \fat \theta^*) }\cdot
	\frac{\pi( \fat \nu^k |\fat \eta, \fat \theta^k) }{\pi( \fat \eta | \fat \nu^k, \fat \theta^k) \pi( \fat \nu^k | \fat \theta^k) }
\end{align*}
\STATE Calculate $\alpha  = \min\left\{ 1, r \right\}$
\STATE Sample $u \sim \mathcal{U}(0,1)$
\IF{$\alpha > u$}
    \STATE $(\fat \nu^{k+1}, \fat \theta^{k+1})= (\fat \nu^*, \fat \theta^*)$
  \ELSIF{$\alpha < u$}
  \STATE $(\fat \nu^{k+1}, \fat \theta^{k+1})  = (\fat \nu^k, \fat \theta^k)$ 
 \ENDIF
\ENSURE $(\fat \nu^{k+1}, \fat \theta^{k+1})$
\end{algorithmic}
\end{algorithm}

\subsection {Spatial prediction} 
\label{Sect:SP}

By using the SPDE spatial model structure, the posterior distribution of all the spatially varying model parameters can be obtained at every regular grid point in $\mathcal G$. The details are as follows.

Let $\fat u ^{[k]}$ be the $k$-th posterior MCMC sample of the spatial effects, for either the location parameter or the scale parameter. The spatial effects $\fat u ^{[k]}$ are located at the vertices of the triangles in the mesh, which do not necessarily coincide with the regular grid points in $\mathcal G$. However, since every regular grid point belongs to some triangle in the mesh, the $k$-th posterior MCMC sample for the spatial effects at the regular grid points can be obtained with a convex linear combination of $\fat u ^{[k]}$, that is,
\begin{equation}
\label{estspat}
	\fat u ^{[k]}_{\mathcal G} = \fat A_{\mathcal G} \fat u ^{[k]}. 
\end{equation}
where the matrix $\fat A_{\mathcal G}$ denotes the linear projection from the vertices of the mesh onto the regular grid points in $\mathcal G$. The term $\fat u ^{[k]}_{\mathcal G}$ then serves as the $k$-th posterior MCMC sample for the spatial effects on the regular grid $\mathcal G$, which is calculated in post calculations after the MCMC run. Therefore, after calculating $\fat u ^{[k]}_{\mathcal G}$ for every iteration $k$, posterior statistics for the spatial effects can be obtained for every point in $\mathcal G$, in particular posterior means and standard deviations. 

The covariates, for both the location and scale parameter, are available by construction at every regular grid point. Thus, the $k$-th posterior MCMC sample for the location parameter can be calculated with
\begin{equation}
\label{muspat}
	\fat \mu^{[k]}_{\mathcal G} = \fat X_\mu \fat \beta^{[k]} + \fat A_{\mathcal G} \fat u_\mu ^{[k]}.
\end{equation}
for every regular grid point. Analogous results hold for the scale parameter. Consequently, the posterior distribution of the $p$-th quantile of the generalized extreme value distribution can be calculated at every regular grid point. The $p$-th quantile function of the generalized extreme value distribution is
\begin{equation}
\label{quantpre}
	q_p(\mu,\tau,\xi) = \mu + \frac{\exp({\tau})}{{\xi}}\left(-\log(p)^{-{\xi}}-1 \right).
\end{equation}
The $k$-th posterior MCMC sample for the $p$-th quantile is then obtained by plugging in the $k$-th posterior MCMC samples for the location, scale and shape parameters. Thus, posterior samples and posterior statistics for the $p$-th quantile are obtained for every point in $\mathcal G$.

\section{Results }
The main objective of this analysis is to obtain spatial predictions for the location and the log-scale parameters and to evaluate the posterior distribution of the 0.95 quantile of maximum precipitation across Iceland on the regular grid $\mathcal G$. In Section 4.1, results from the MCMC convergence diagnostics are briefly summarized while in Section 4.2 tables and figures of posterior estimates of the model parameters are given and discussed. In Section 4.3 figures of spatial predictions are shown and discussed.

\subsection{Convergence diagnostics}
The following results are based on four MCMC chains from the sampler in Section \ref{SplitSampler}. Each chain is calculated with 35000 iterations where 10000 iterations were burned in. Runtime, on a modern desktop (Ivy Bridge Intel Core i7-3770K, 16GB RAM and solid state hard drive), was approximately four hours for each data set. All the calculations were done using $\mathtt R$.

\begin{figure}[htbp]
   \centering
   \includegraphics[width=1 \textwidth]{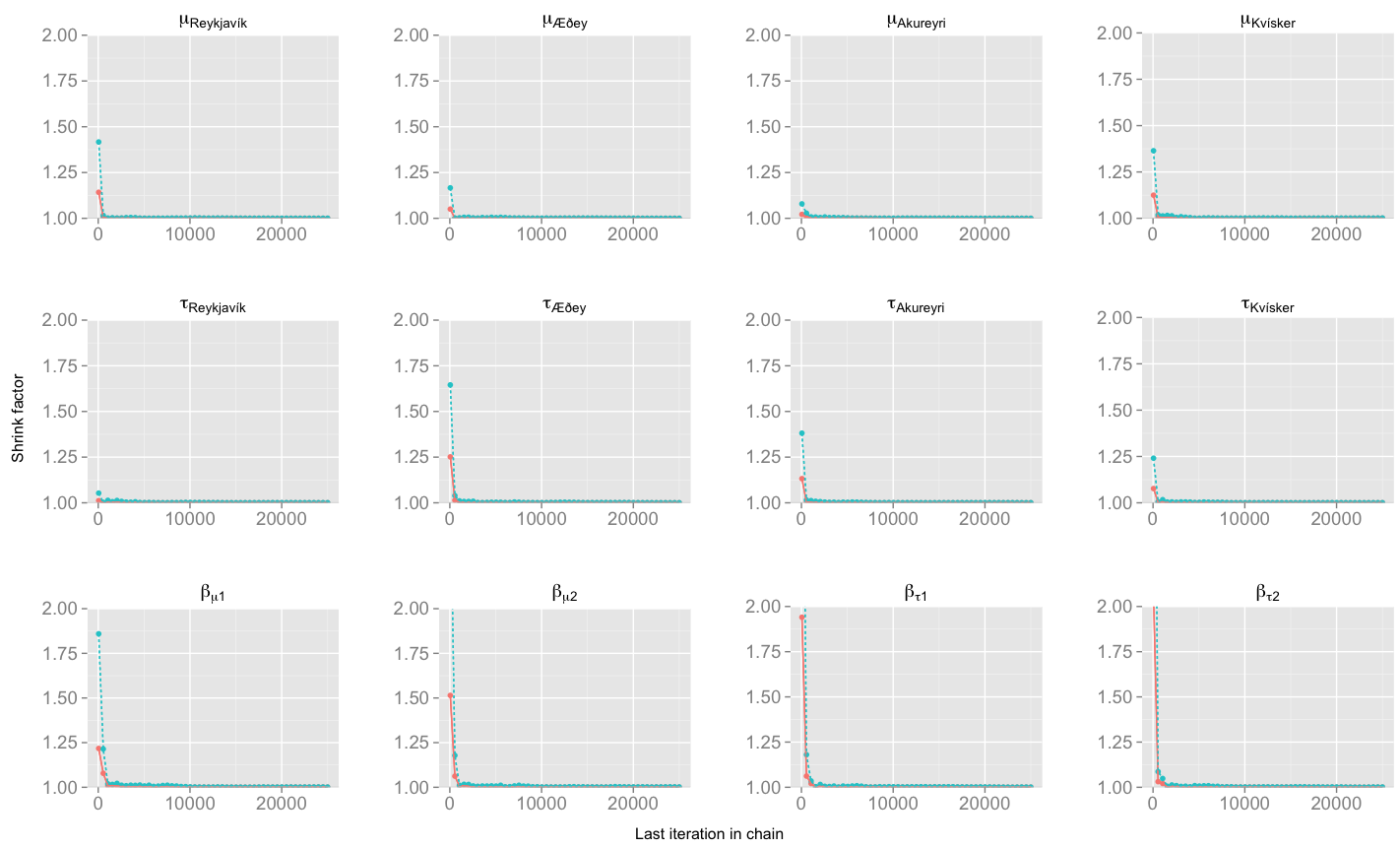}
   \caption{The first two rows show Gelman--Rubin statistics as a function of iterations after burn-in, for $\fat \mu$ and $\fat \tau$, respectively,  at Reykjavík, Æðey, Akureyri and Kvísker, based on the uncorrected data set. The third row shows a  Gelman--Rubin plot for the coefficients of the covariates.}
   \label{GR_latent}
\end{figure}

\begin{figure}[htbp]
   \centering
   \includegraphics[width=1 \textwidth]{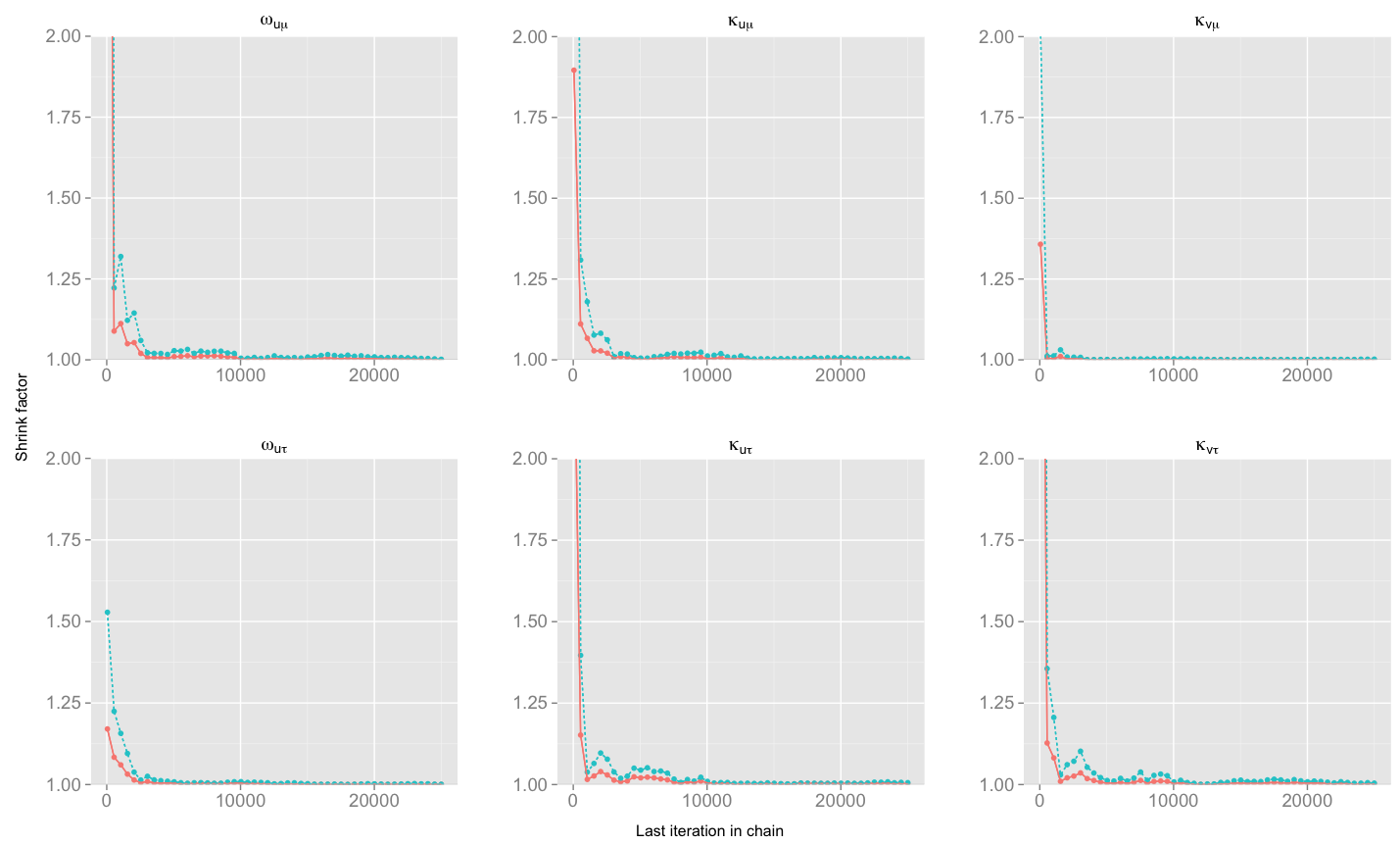}
   \caption{The figure shows Gelman--Rubin statistics as a function of iterations after burn-in, for all the hyper parameters $\kappa_{u \mu}$, $\omega_{u \mu}$, $\kappa_{u \tau}$, $\omega_{u \tau}$, $\kappa_{v \mu}$ and $\kappa_{v \tau}$ based on the uncorrected data set}
   \label{GR_hyper}
\end{figure}

Gelman--Rubin statistics \citep{brooks1998general} were calculated for all model parameters based on the four MCMC chains for both data sets. The Gelman--Rubin statistics for all the model parameters in both data sets, were evaluated as approximately 1, indicating that all the MCMC chains have converged in the mean. Figure \ref{GR_latent} shows Gelman--Rubin plots for the location parameter $\fat \mu$ and log-scale parameter $\fat \tau$ in Reykjavík, Æðey, Akureyri and Kvísker; and for the coefficients of the covariates $\fat \beta_\mu$ and $\fat \beta_\tau$. Furthermore, Figure \ref{GR_hyper} shows Gelman--Rubin plots for all the hyperparameters. Both figures are based on the uncorrected data set. Similar results hold for the corrected data set. Moreover,  Figures \ref{GR_latent} and \ref{GR_hyper} indicate that the sampler converges after 2000 iterations after burn-in. 

\begin{figure}[htbp]
   \centering
   \includegraphics[width=1 \textwidth]{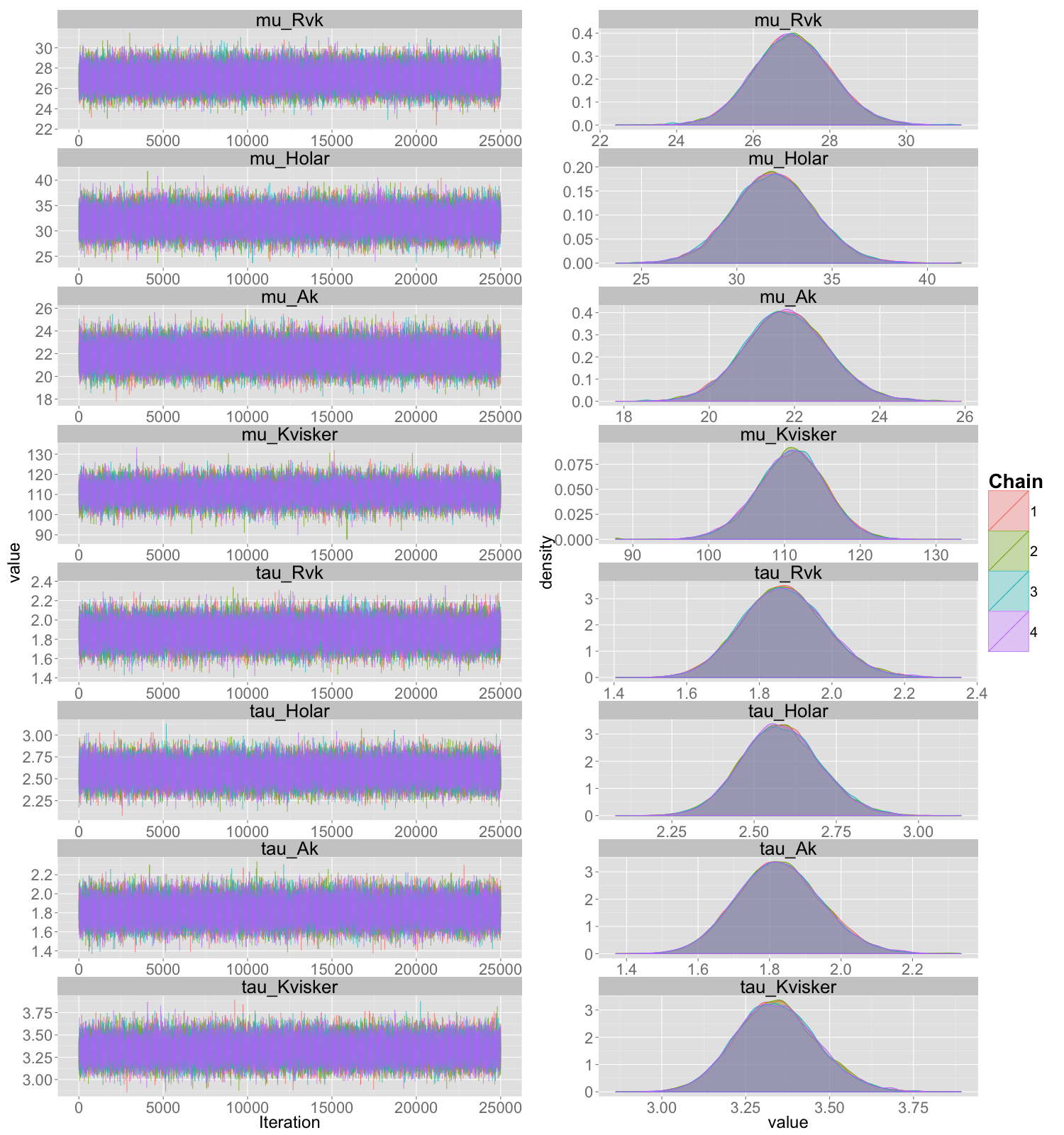}
   \caption{Posterior trace and density plots of $\mu$ and $\tau$ at Reykjavík, Æðey, Akureyri and Kvísker, based on the uncorrected data set}
   \label{fig:DigLatent1}
\end{figure}

\begin{figure}[htbp]
   \centering
   \includegraphics[width=1 \textwidth]{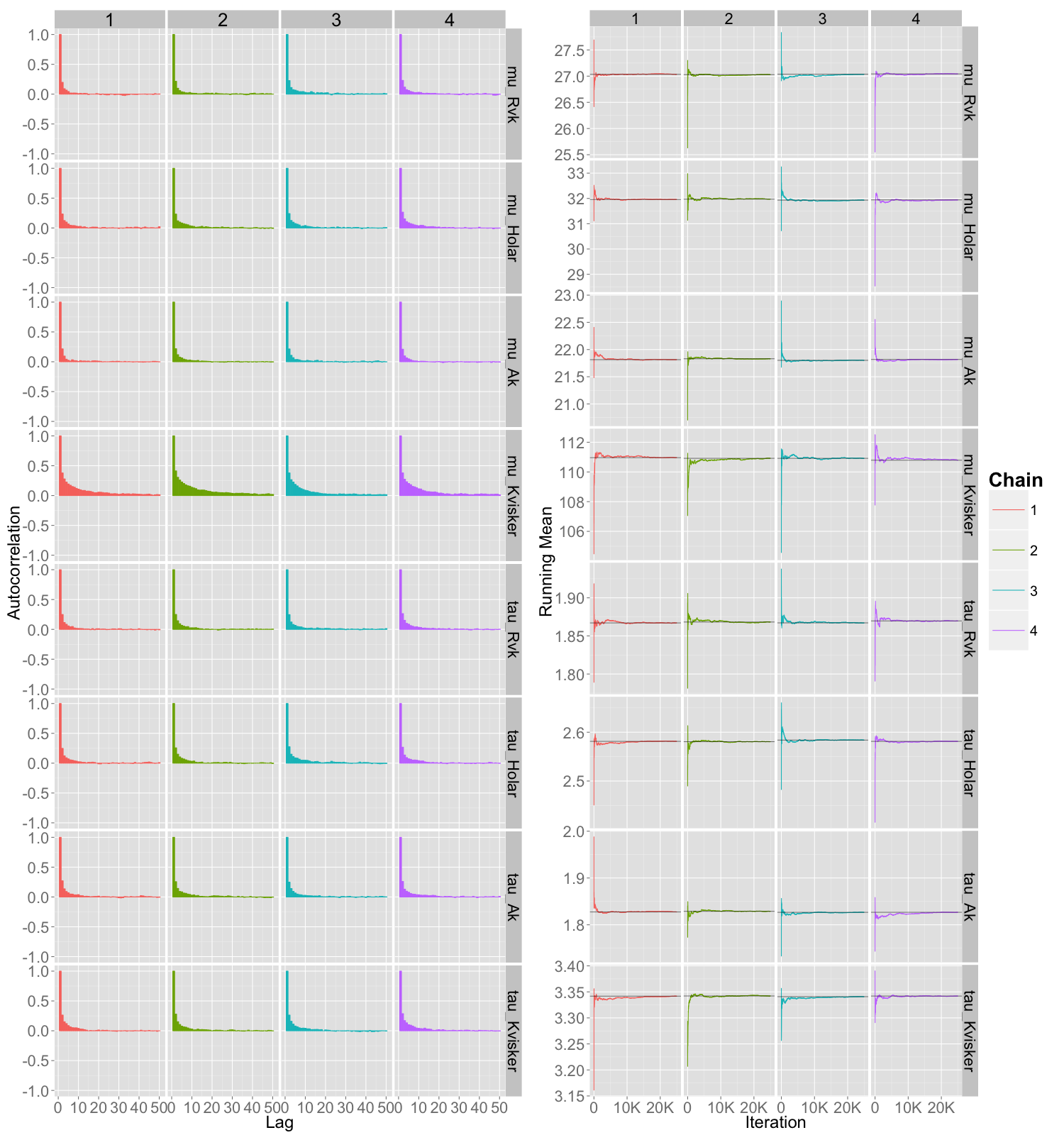}
	\captionof{figure}{Auto correlation and running mean plots of $\mu$ and $\tau$ at Reykjavík, Æðey, Akureyri and Kvísker, based on the uncorrected data set}
	\label{fig:DigLatent2} 
\end{figure}

\begin{figure}[htbp]
   \centering
   \includegraphics[width=1 \textwidth]{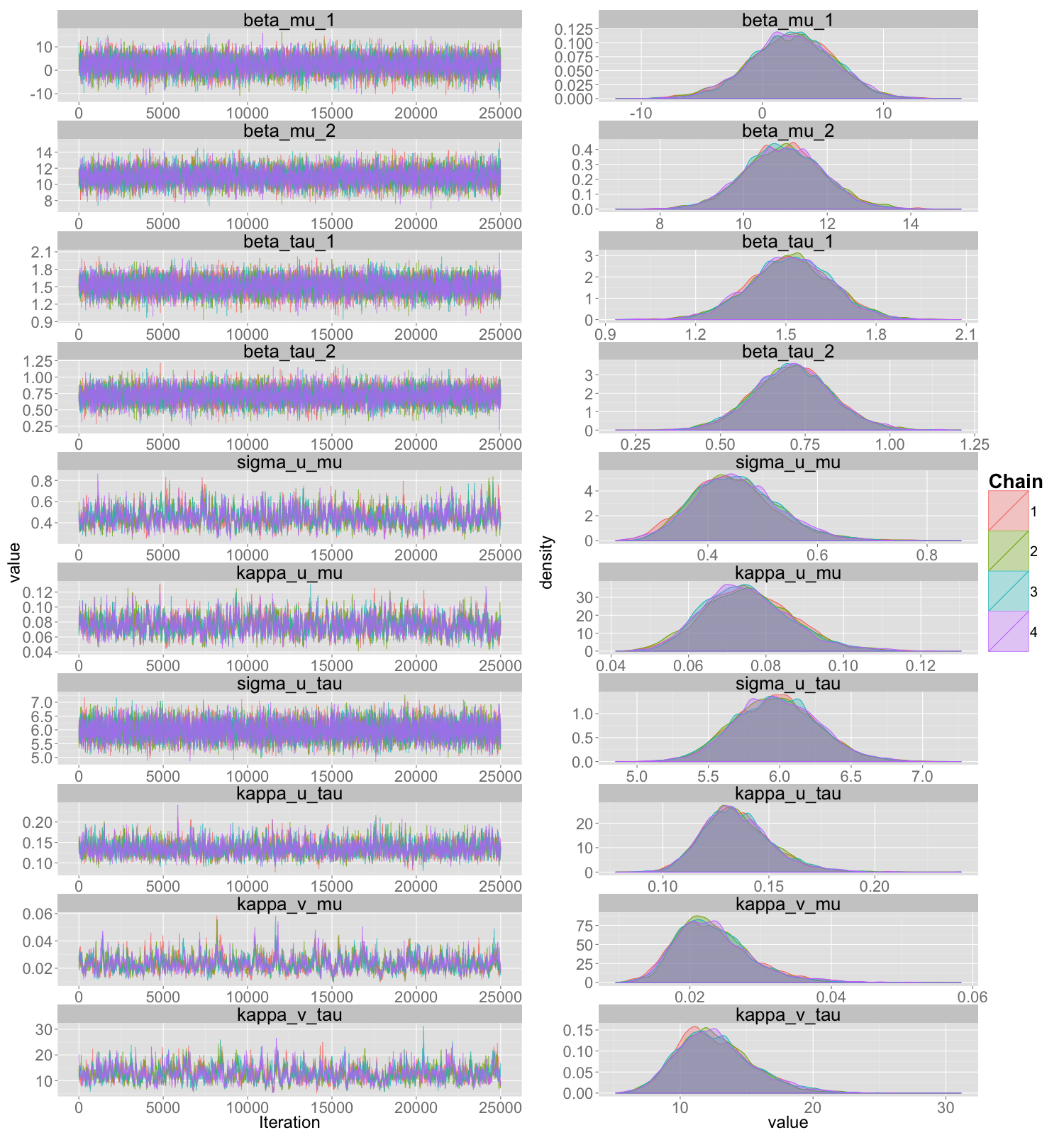}
   \caption{Posterior trace and density plots of the coefficients of the covariates $\fat \beta_\mu$ and $\fat \beta_\tau$; and of the all hyperparameters, based on the uncorrected data set.}
   \label{fig:DigHyp1}
\end{figure}

\begin{figure}[htbp]
   \centering
  \includegraphics[width=1 \textwidth]{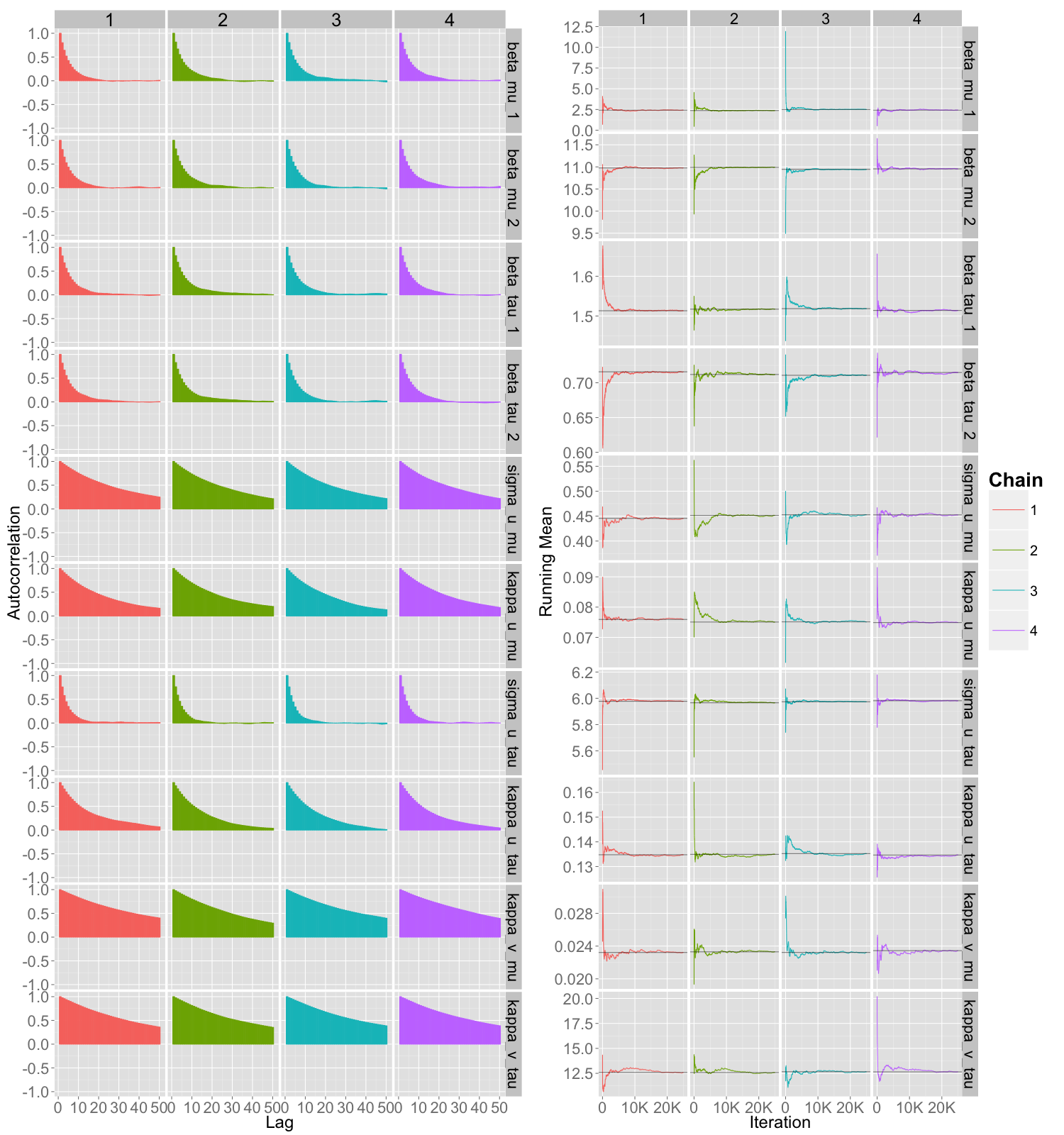}
   \caption{Auto correlation and running mean plots of the coefficients of the covariates $\fat \beta_\mu$ and $\fat \beta_\tau$; and of the all hyperparameters, based on the uncorrected data set.}
   \label{fig:DigHyp2}
\end{figure}

Others convergence diagnostics plots can be seen in the Appendix for the same set of parameters, where the four MCMC chains were evaluated. Figure \ref{fig:DigLatent1} shows trace and density plots of the location and log-scale parameters for the same observational sites as above;  autocorrelation and running mean plots can be seen in Figure \ref{fig:DigLatent2}. Figure \ref{fig:DigHyp1} and Figure \ref{fig:DigHyp2} show the same two convergence diagnostics plots for the coefficients of the covariates  and all the hyperparameters The MCMC chains for the latent parameters and the coefficients of the covariates  exhibit neglectable autocorrelation after lag 10. Samples of the hyperparameters show some autocorrelation, but within an acceptable range as they are highly correlated in the posterior.

These convergence diagnostics results demonstrate the sampler's efficiency, and strengthen the claim that the sampler converges quickly. We state, with fair amounts of confidence, that the sampler is highly efficient for this model.

\subsection{Posterior estimates}
\label{PE}
\begin{figure}[htp]
	\centering
	\begin{subfigure}[b]{0.47 \textwidth}
    	\includegraphics[width=\textwidth,]{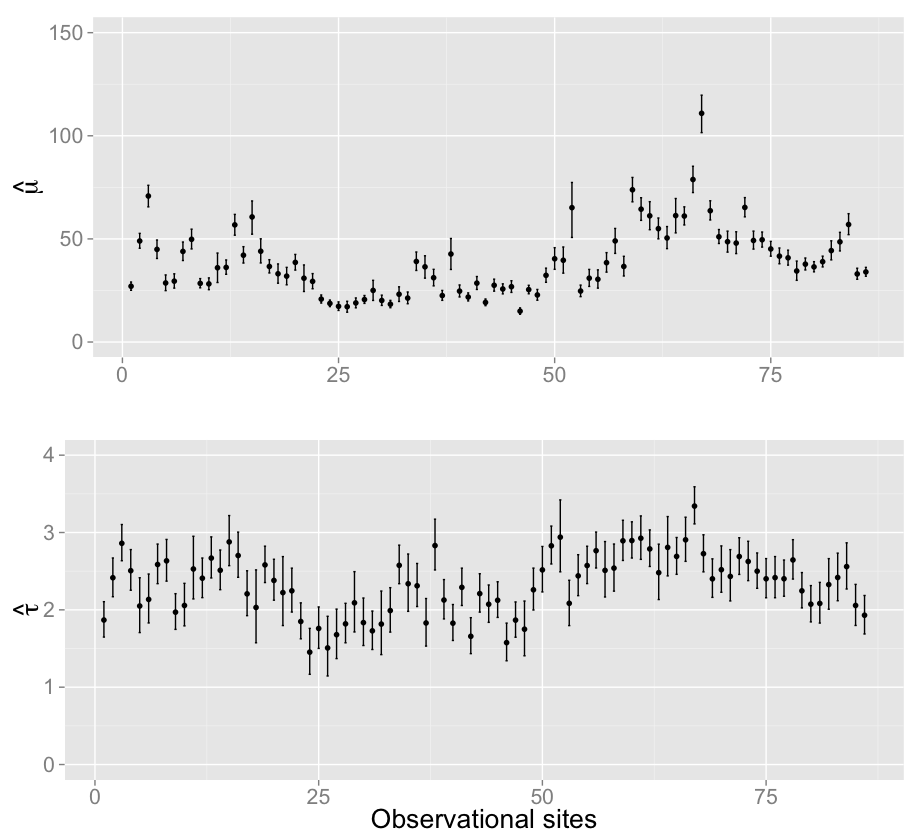}
    	\caption{Uncorrected data set}
  	\end{subfigure}
	\hspace{0.5cm}
  	\begin{subfigure}[b]{0.47 \textwidth}
    	\includegraphics[width=\textwidth]{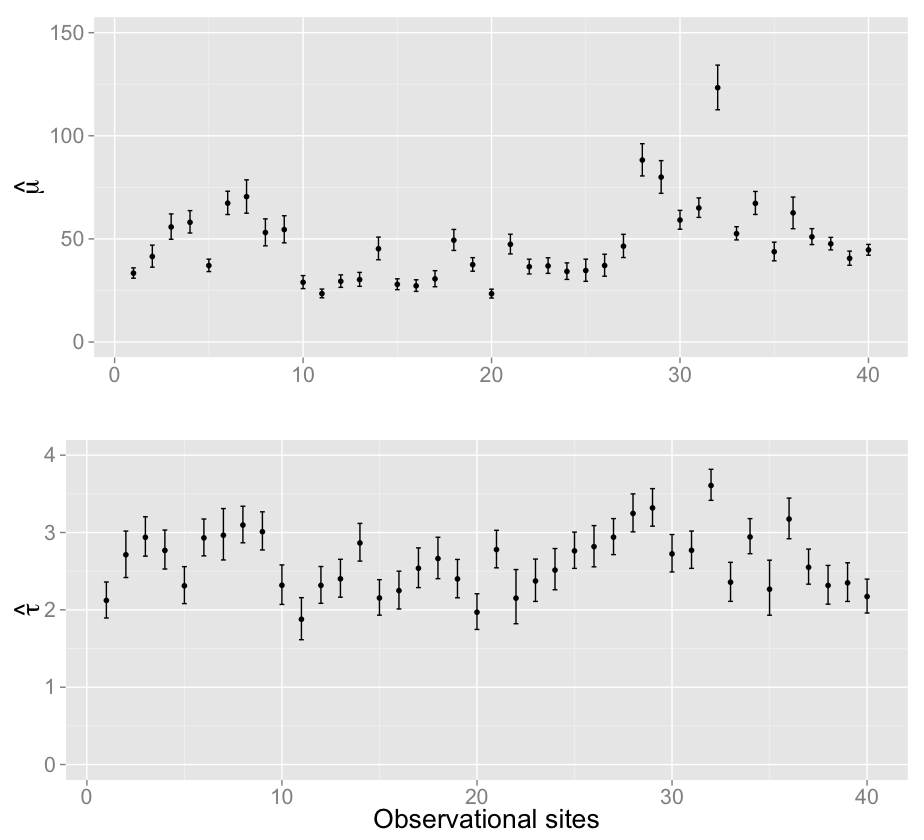}
    	\caption{Corrected data set}
  \end{subfigure}
  \caption{The first and second row show posterior estimates of $\fat \mu$ and $\fat \tau$, respectively, for each observational site}
  \label{supernal}
\end{figure}

The following posterior estimates are based on the four MCMC chains after burn-in. Figure \ref{supernal} shows the posterior mean with 95\% posterior interval for both the location and log-scale parameters for each observational site. The sites are placed on the $x$-axis as follows. The leftmost site on the $x$-axis is in Reykjavík. The rest of sites are places on the $x$-axis corresponding to a clockwise labeling across the sites shown in Figure \ref{FIG:obs}. The results in Figure \ref{supernal} suggest that both the average and standard deviation of extreme precipitation is the highest and the south-eastern part in Iceland, and the lowest in the northern part.

Table \ref{ParaTable} shows the posterior mean, posterior standard deviation and posterior $0.025$ and $0.975$ quantile estimates for the non-spatially varying model parameters. The posterior results are similar for all the parameters between the two data sets, except for the covariate coefficients $\beta_{\mu 1}$ and $\beta_{\mu 2}$. That is  to be expected, as every measurement in the corrected data set is greater or equal to the corresponding measurement in the uncorrected data set, by construction. The following results are discussed for the uncorrected data set. Analogous results and interpretation hold for the corrected data set.

\begin{table}[htb]
\centering
\begin{tabular}{|l|rrrr@{\hskip 0.2in}|rrrr@{\hskip 0.2in}|}
  \hline
  Parameter & \multicolumn{3}{ l }{Uncorrected data} && \multicolumn{3}{ l }{Corrected data}  &\\ 
 &  0.025 & mean & 0.975 & sd &  0.025 & mean & 0.975 & sd \\ 
  \hline
$\beta_{\mu 1}$ & -4.65 & 2.41 & 8.96 & 3.46 & 2.53 & 11.76 & 20.82 & 4.66 \\ 
  $\beta_{\mu 2}$ & 9.14 & 10.97 & 12.87 & 0.95 & 9.04 & 11.59 & 14.24 & 1.33 \\ 
  $\beta_{\tau 1}$ & 1.24 & 1.52 & 1.79 & 0.14 & 1.68 & 2.10 & 2.52 & 0.21 \\ 
  $\beta_{\tau 2}$ & 0.49 & 0.71 & 0.94 & 0.12 & 0.13 & 0.50 & 0.87 & 0.19 \\ 
  $\xi$ & 0.06 & 0.09 & 0.12 & 0.02 & 0.08 & 0.12 & 0.15 & 0.02 \\ 
  $\omega_{u \mu}$ & 0.31 & 0.45 & 0.62 & 0.08 & 0.28 & 0.40 & 0.55 & 0.07 \\ 
  $\kappa_{u \mu}$ & 0.05 & 0.08 & 0.10 & 0.01 & 0.05 & 0.07 & 0.10 & 0.01 \\ 
  $\omega_{u \tau}$ & 5.42 & 5.97 & 6.57 & 0.29 & 5.37 & 5.93 & 6.53 & 0.29 \\ 
  $\kappa_{u \tau}$ & 0.11 & 0.13 & 0.17 & 0.02 & 0.09 & 0.12 & 0.16 & 0.02 \\ 
  $\kappa_{v \mu}$ & 0.01 & 0.02 & 0.04 & 0.01 & 0.01 & 0.02 & 0.03 & 0.01 \\ 
  $\kappa_{v \tau}$ & 7.85 & 12.61 & 19.05 & 2.87 & 4.04 & 7.10 & 11.61 & 1.94 \\ 
   \hline
\end{tabular}
\caption{Posterior $0.025$ quantile, mean, $0.975$ quantile and standard deviation of the above model parameters}
\label{ParaTable}
\end{table}

The posterior density for $\beta_{\mu 2}$ suggests that the effects of the constructed meteorological covariate is close to $11$ and it is positive as its 95\% posterior interval is above zero. Assuming the simulated means approximate the observed means well enough, these results indicate that the location parameter describing the extreme precipitation is roughly 11 times the corresponding annual mean in case of the uncorrected data. 

The posterior density for $\beta_{\tau 1}$ and  $\beta_{\tau 2}$ yield a point estimates close to $1.5$ and $0.7$, respectively, and it reveals that they are both positive as their 95\% posterior intervals are above zero. That means that the relationship between the simulated means and the scale parameter of the generalized extreme value distribution can be roughly summarized as
\[	
	\hat{\sigma}_i = e^{\hat{\beta}_1}  \bar x^{* \hat \beta_2}_i \approx 4.6 \cdot   \bar x^{* 0.7}_i 
\]

The estimate for the shape parameter $0.09$ and its posterior interval is approximately $(0.06, 0.12)$. That indicates that the proposed generalized extreme values distributions does not have an upper bound and its first and second moments are finite.

The posterior estimates of the parameters $\kappa_{u \mu}$ and $\kappa_{u \tau}$, suggest that the correlation between two points in space is near 0.1 at a roughly 40 km distance for the location parameter, and roughly 25 km distance for the scale parameter. The posterior estimates of the parameters $\omega_{u \mu}$ and $\omega_{u \tau}$, indicate that the marginal standard deviation of $\fat u_\mu$ is approximately 8 and 0.4 for $\fat u_\tau$, which is the spatial effect for the scale parameter on a logarithmic scale.

The posterior estimates of $\kappa_{v \mu}$ indicate that the posterior standard deviation of the unstructured random effect $\fat v_\mu$ is approximately 7. Which in turn suggests that there is some variation left in the data that is unexplained by the covariates and the spatial model. That behavior might be due to the fact that some observational sites, like Kvísker, are located in areas with high topographical variation that yield a non-stationary behavior in the spatial field corresponding to the location parameter.  Similar behavior is observed for the scale parameter, as the posterior estimates of $\kappa_{v \tau}$ indicate that the posterior standard deviation of the unstructured random effect $\fat v_\tau$ is approximately 0.28 on a logarithmic scale.

\begin{figure}[htp]
	\centering
	\begin{subfigure}[b]{0.47 \textwidth}
    	\includegraphics[width=\textwidth,]{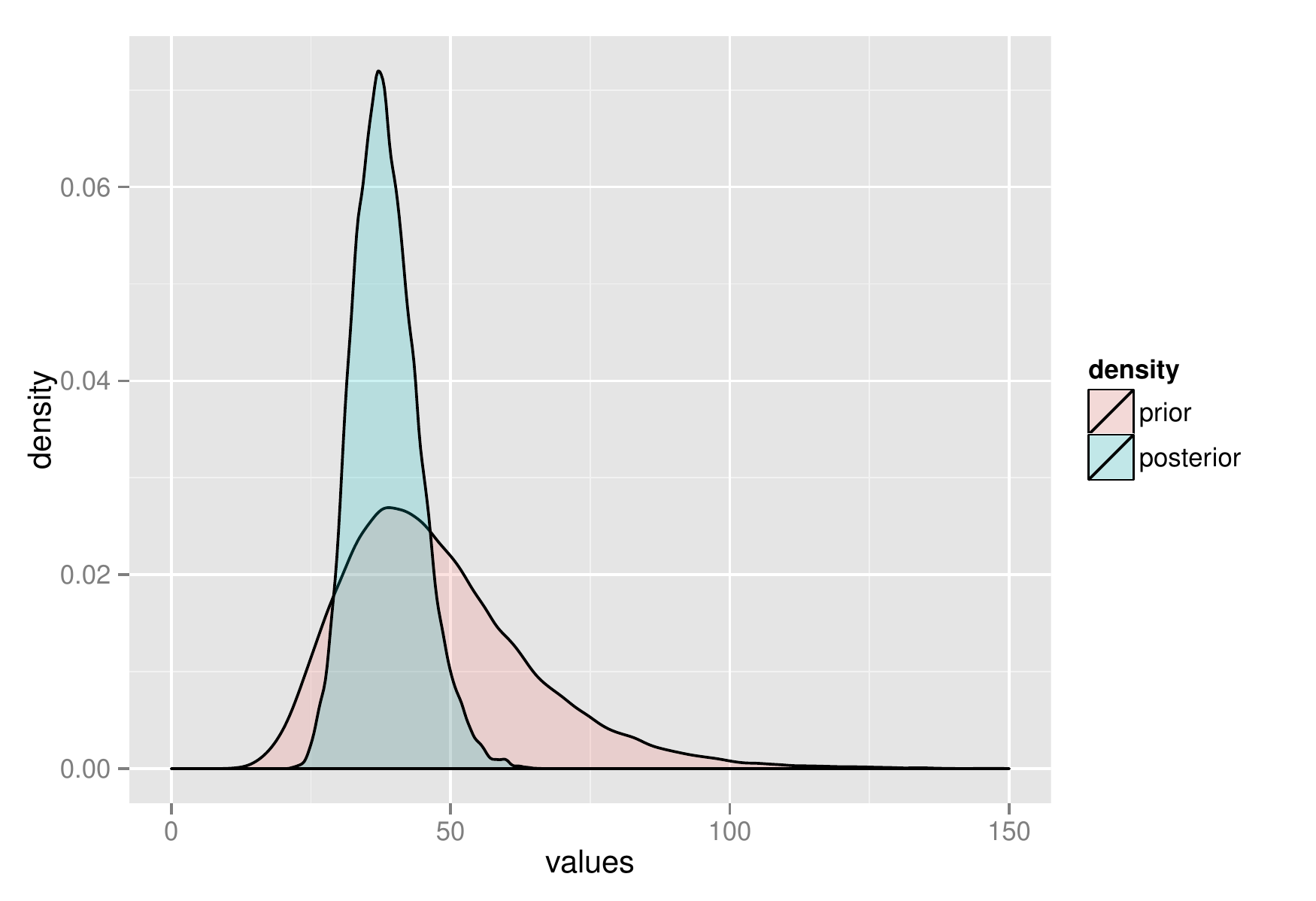}
    	\caption{Range of $\fat u_\mu$}
	 \label{PP_RangeEta}
  	\end{subfigure}
	\hspace{0.5cm}
  	\begin{subfigure}[b]{0.47 \textwidth}
    	\includegraphics[width=\textwidth]{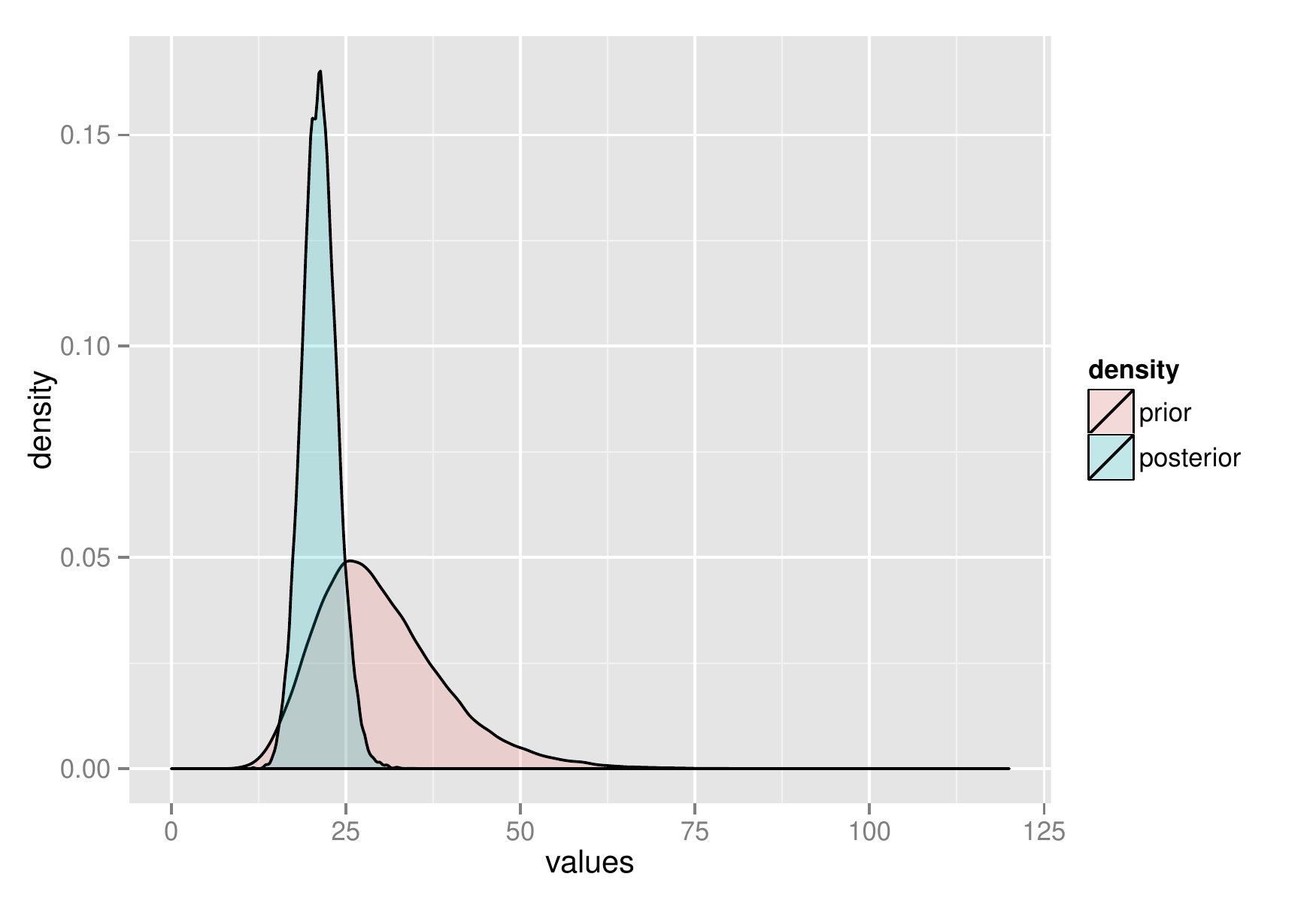}
    	\caption{Range of $\fat u_\tau$}
	 \label{PP_RangeTau}
  	\end{subfigure}
	
	\begin{subfigure}[b]{0.47 \textwidth}
    	\includegraphics[width=\textwidth,]{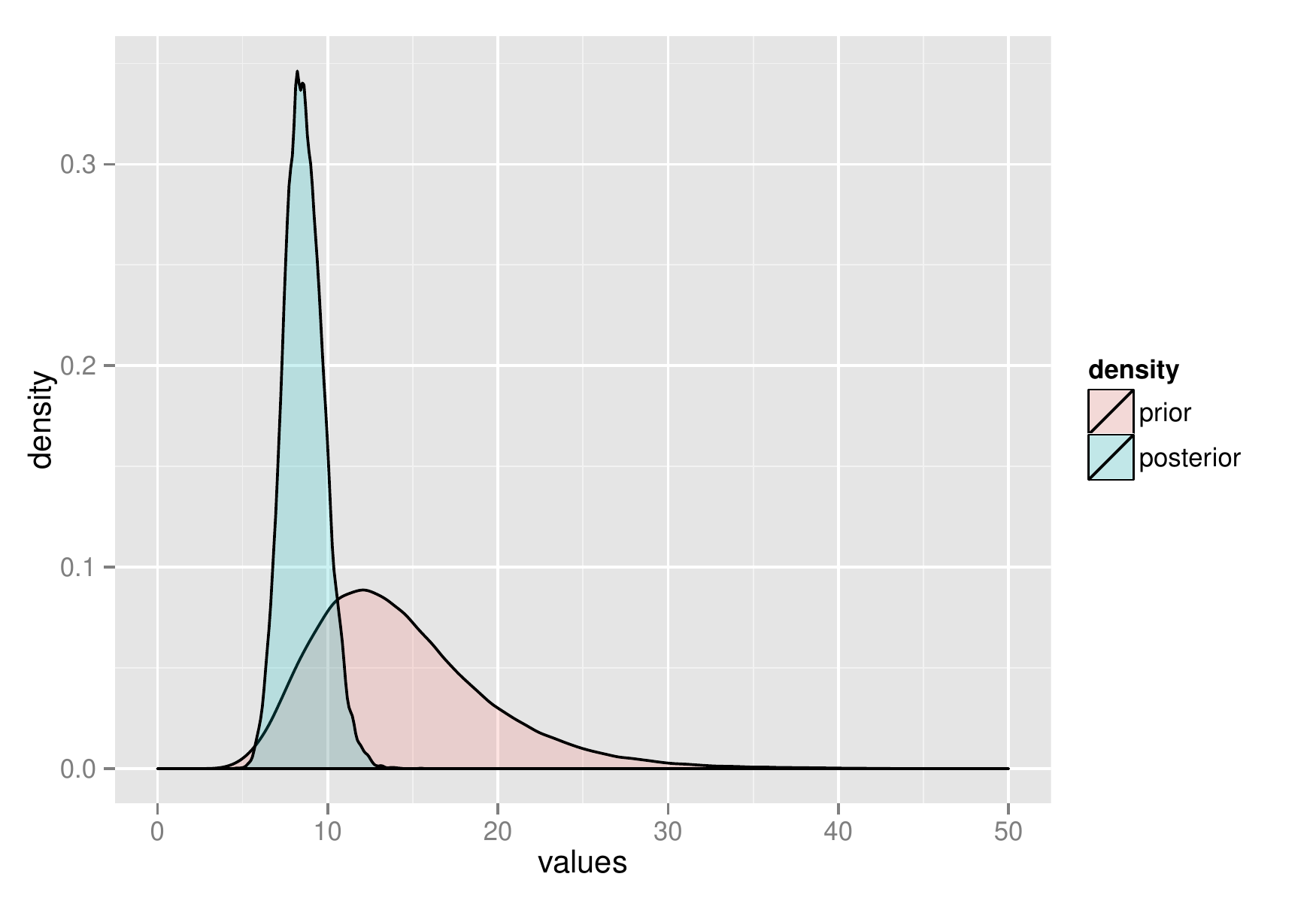}
    	\caption{Standard deviation of $\fat u_\mu$}
	 \label{PostPrior_SdEta}
  	\end{subfigure}
	\hspace{0.5cm}
  	\begin{subfigure}[b]{0.47 \textwidth}
    	\includegraphics[width=\textwidth]{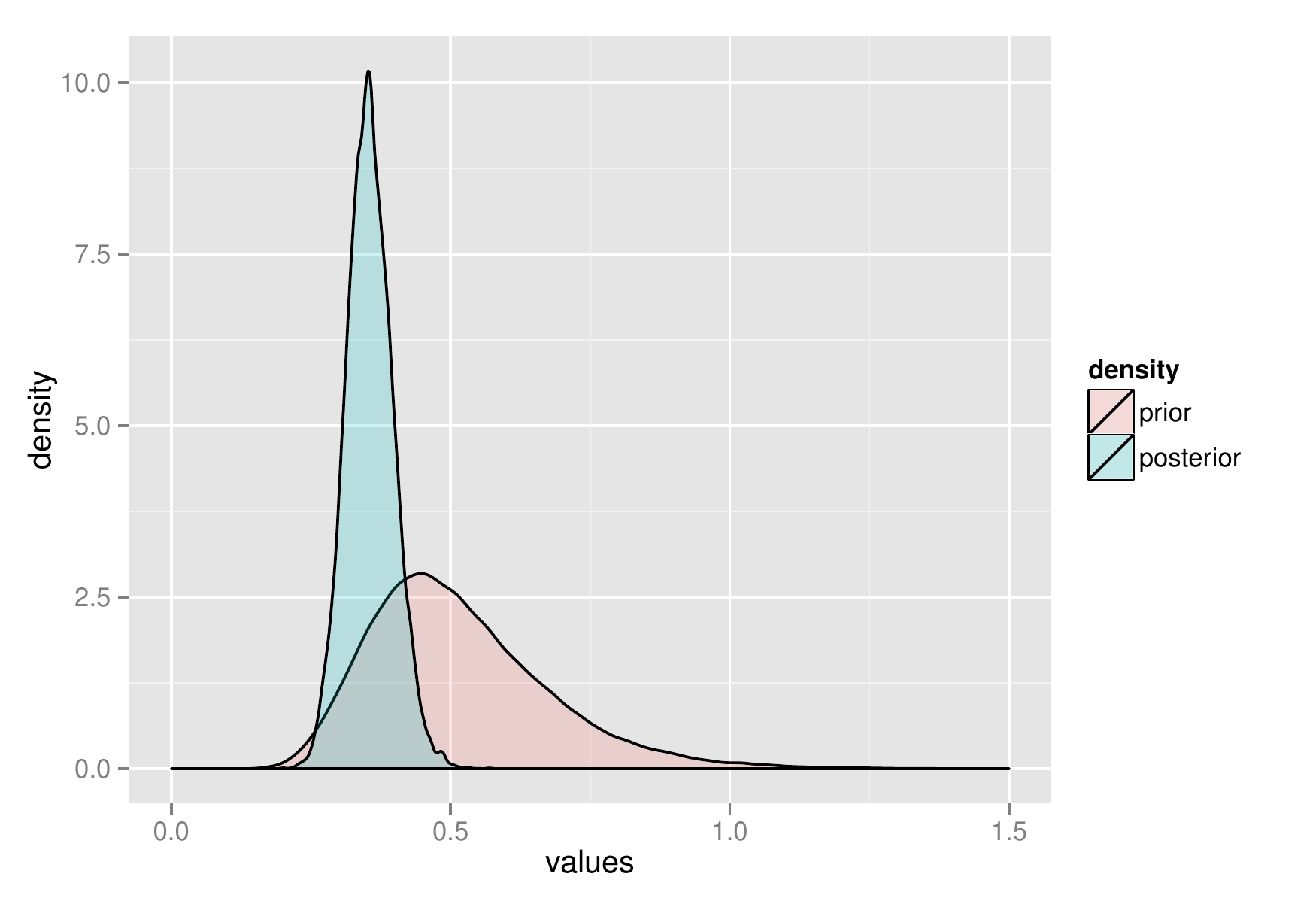}
    	\caption{Standard deviation of $\fat u_\tau$}
	 \label{PostPrior_SdTau}
  	\end{subfigure}
	
	\begin{subfigure}[b]{0.47 \textwidth}
    	\includegraphics[width=\textwidth,]{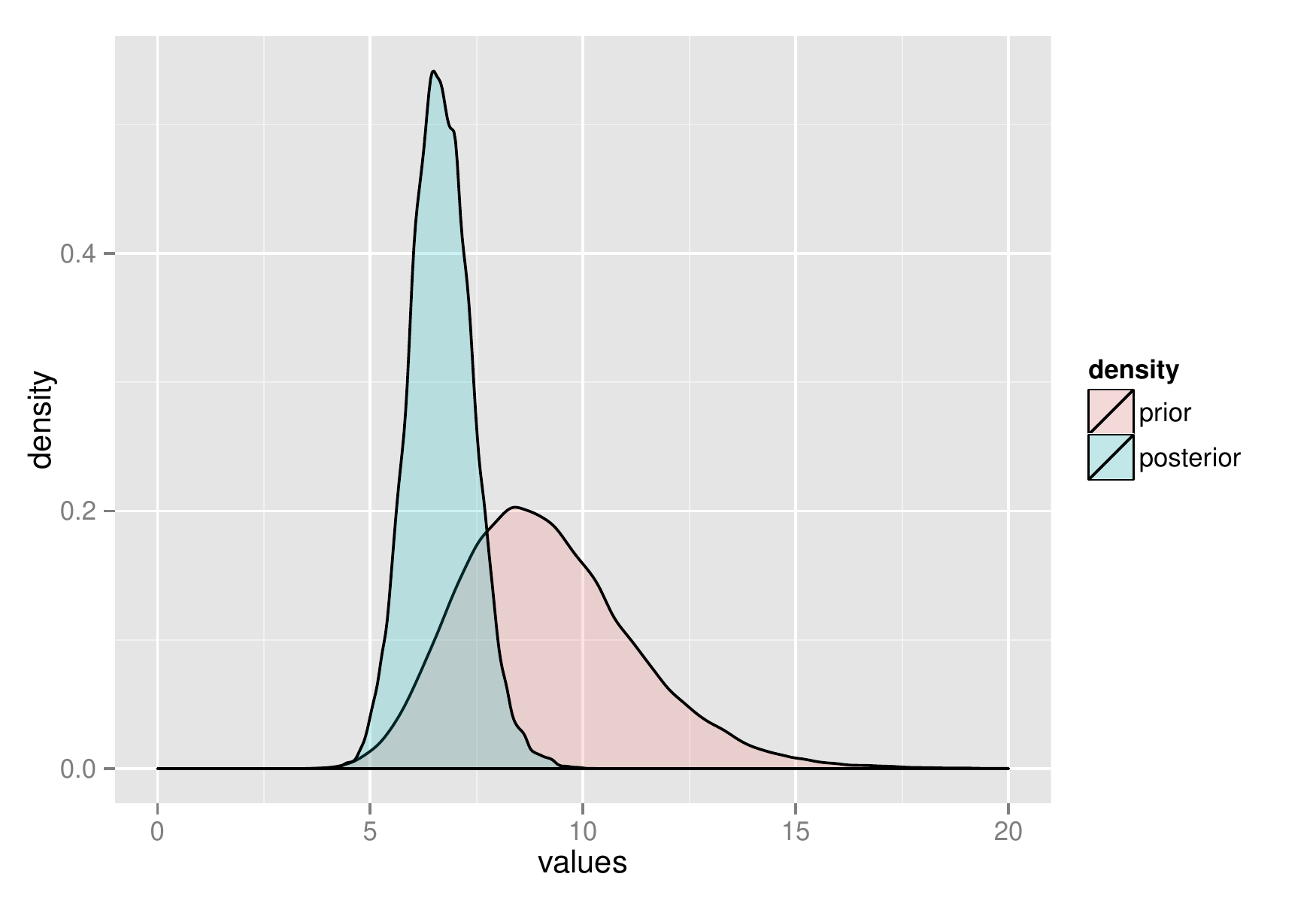}
    	\caption{Standard deviation of $\fat v_\mu$}
	\label{PostPriorVloc}
  	\end{subfigure}
	\hspace{0.5cm}
  	\begin{subfigure}[b]{0.47 \textwidth}
    	\includegraphics[width=\textwidth]{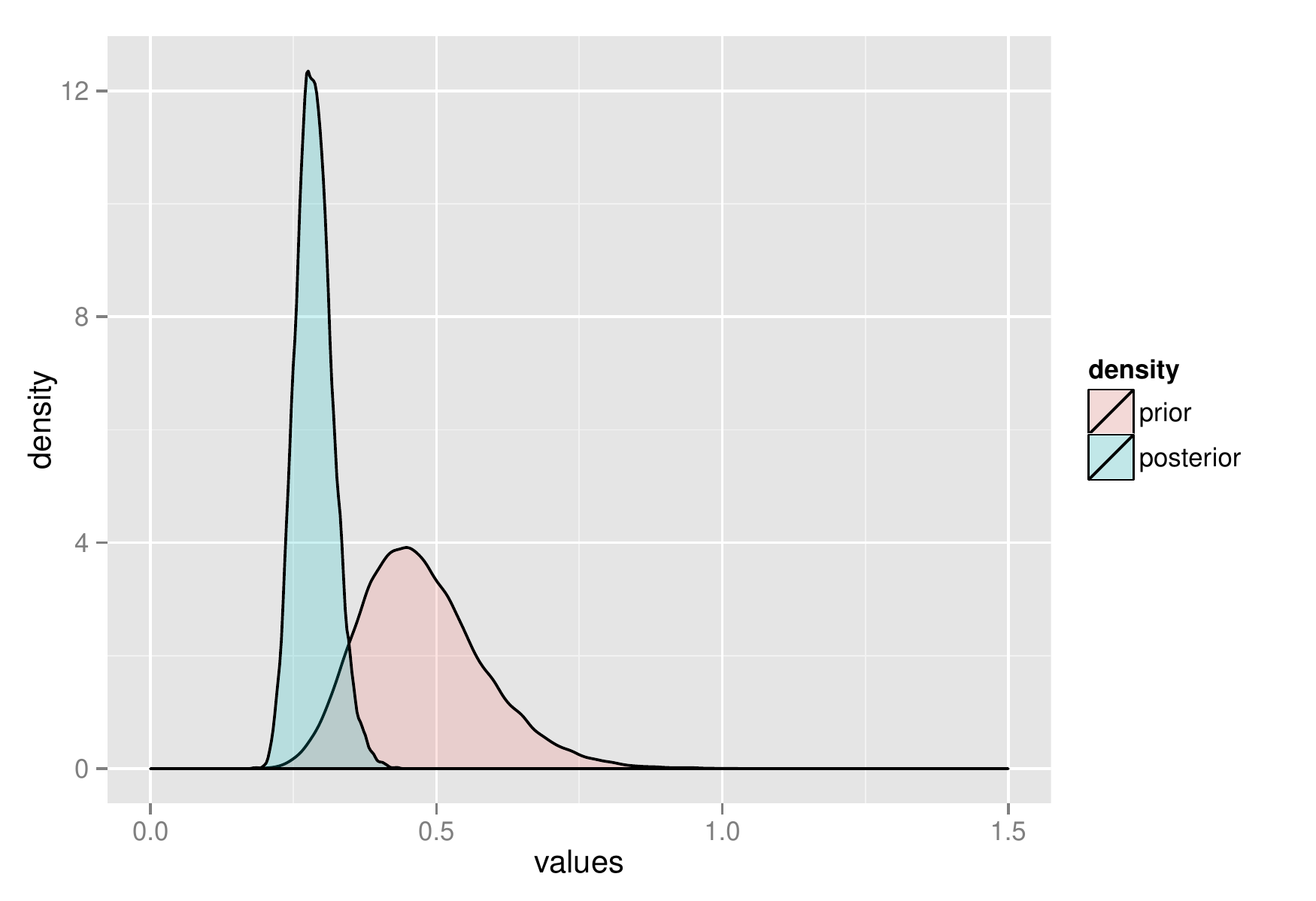}
    	\caption{Standard deviation of $\fat v_\tau$}
	\label{PostPrior_VTau}
  	\end{subfigure}
	
  \caption{Prior distribution for all the hyperparameters and corresponding posterior distributions based on the uncorrected data set.}
  \label{PostPrior}
\end{figure}

In Figure \ref{PostPrior} prior distributions for all the hyperparameters are shown, along with corresponding posterior distributions based on the uncorrected data set. 

\subsection{Model evaluation}

\begin{figure}[h!]
	\centering
	\begin{subfigure}[b]{0.47 \textwidth}
    	\includegraphics[width=\textwidth,]{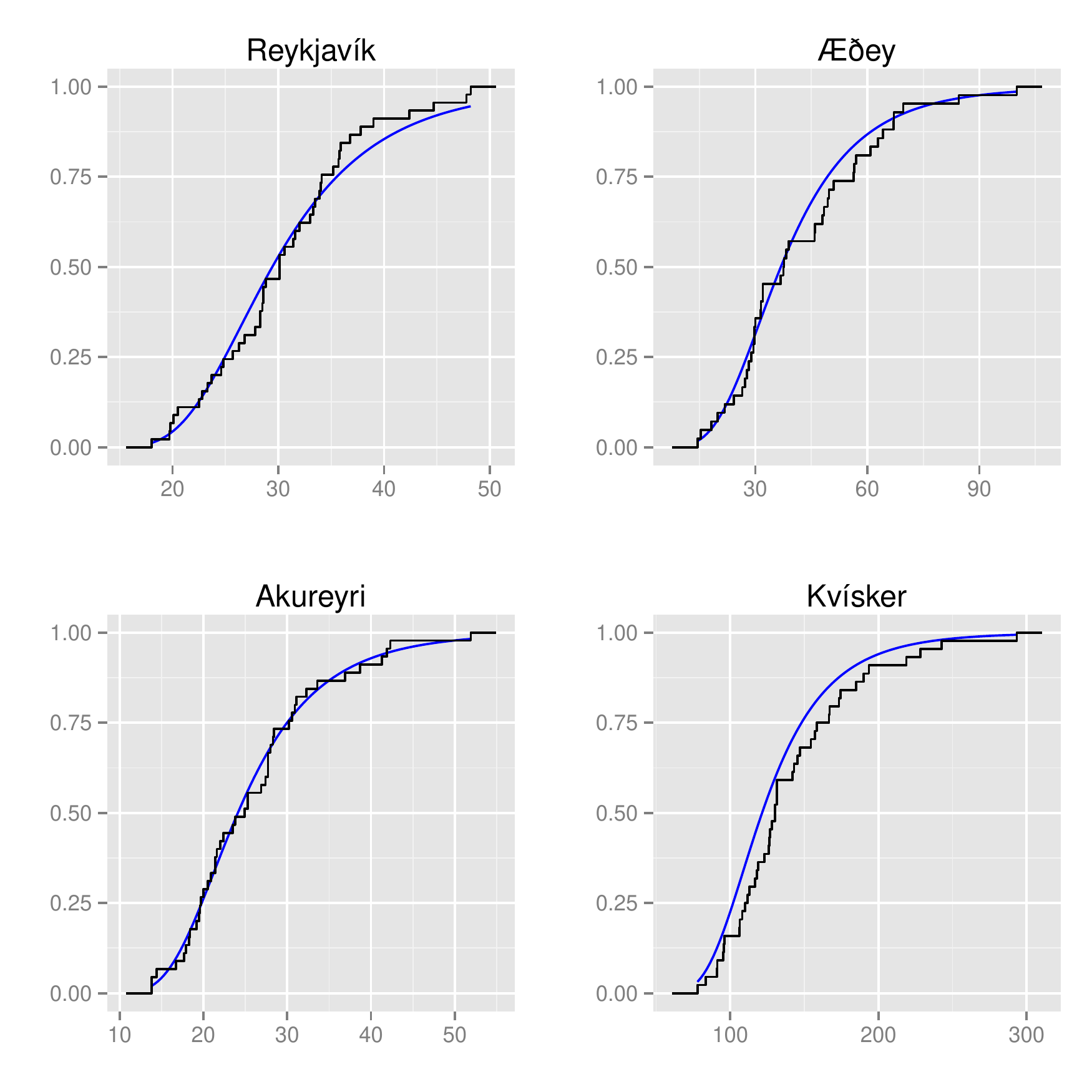}
    	\caption{Uncorrected data set}
    	\label{FIG:CDF1}
  	\end{subfigure}
	\hspace{0.5cm}
  	\begin{subfigure}[b]{0.47 \textwidth}
    	\includegraphics[width=\textwidth]{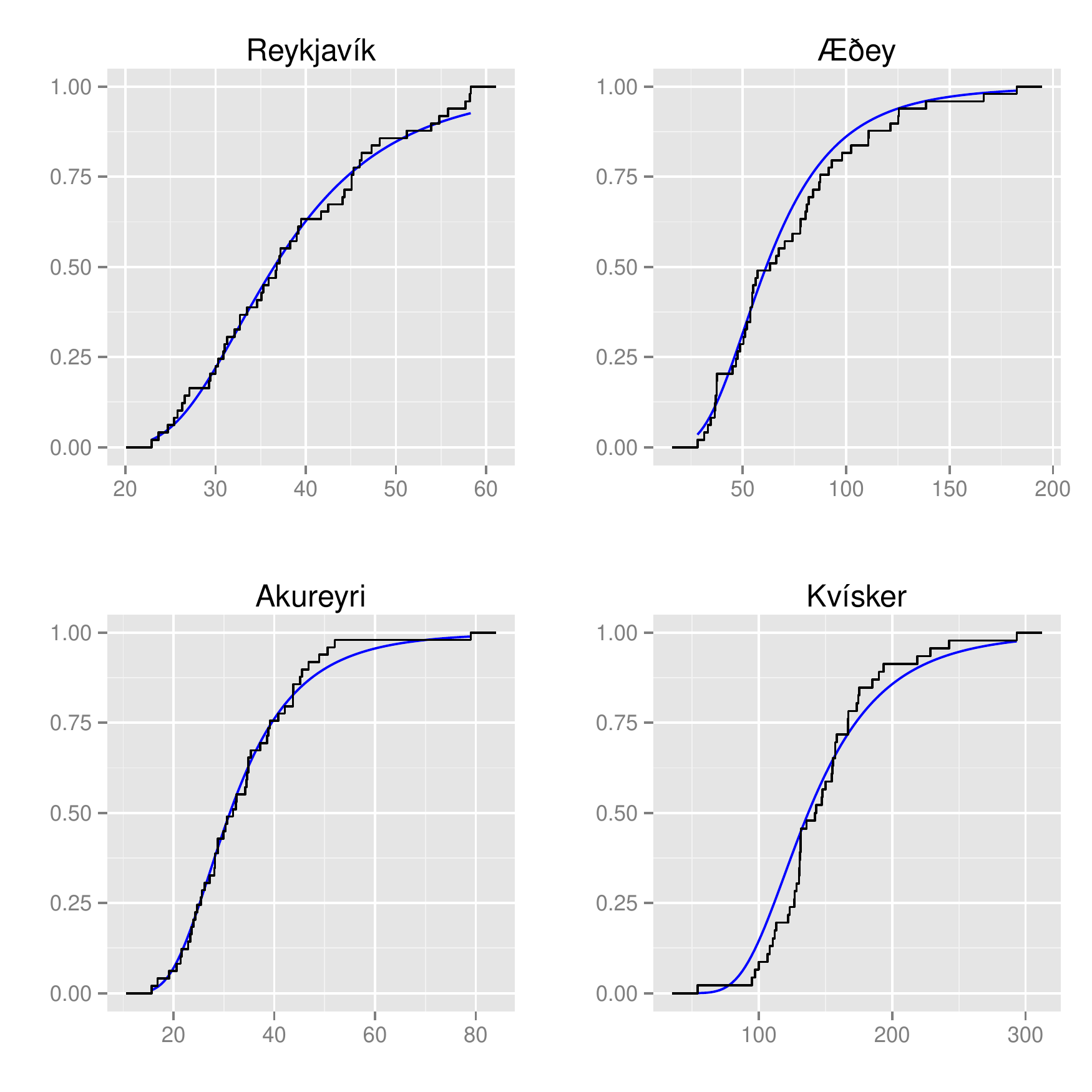}
    	\caption{Corrected data set}
    	\label{FIG:CDF2}
  \end{subfigure}
  \caption{Empirical (black) and model cumulative (blue) distribution functions.}
  \label{CDFin}
\end{figure}

In Figure \ref{CDFin} empirical cumulative distributions for the observational sites Reykjavík, Æðey, Akureyri and Kvísker  are compared to corresponding  posterior cumulative distributions functions uncorrected data set. The results indicate that the model describes the data well. The observation at Kvísker deviate the most from the model in the uncorrected data set, while that behavior is not seen in the corrected data set. This might be due to that fact that there are observational sites close to Kvísker in the uncorrected data set that both show much lower extremal observations and are not included in the corrected data set.

An overall time effect for location parameter was evaluated based on the fitted values at each observational site. That is, let $m_{it} = y_{it} - \text{E}(y_{it} )$, where $\text{E}(y_{it} )$ is the excepted value of the g.e.v. model with parameter values based on posterior means. A likelihood ratio test was applied to a model for the $m$'s with an overall time effect against a model for the $m$'s without a time effect. The results indicated no significant difference between the two models (significance level $0.05$). Thus, there was no reason to add an overall time effect to the proposed model.

\subsection{Spatial predictions}
The following figures show the spatial predictions for the spatially varying model parameters, based on the methods from Section \ref{Sect:SP}. Every figure in this section on the left panels are based on the uncorrected data set, and the figures on the right hand side are based on the corrected data set.

\begin{figure}[htp]
	\centering
	\begin{subfigure}[b]{0.495 \textwidth}
    	\includegraphics[width=\textwidth, trim =  15mm 25mm 30mm 20mm, clip]{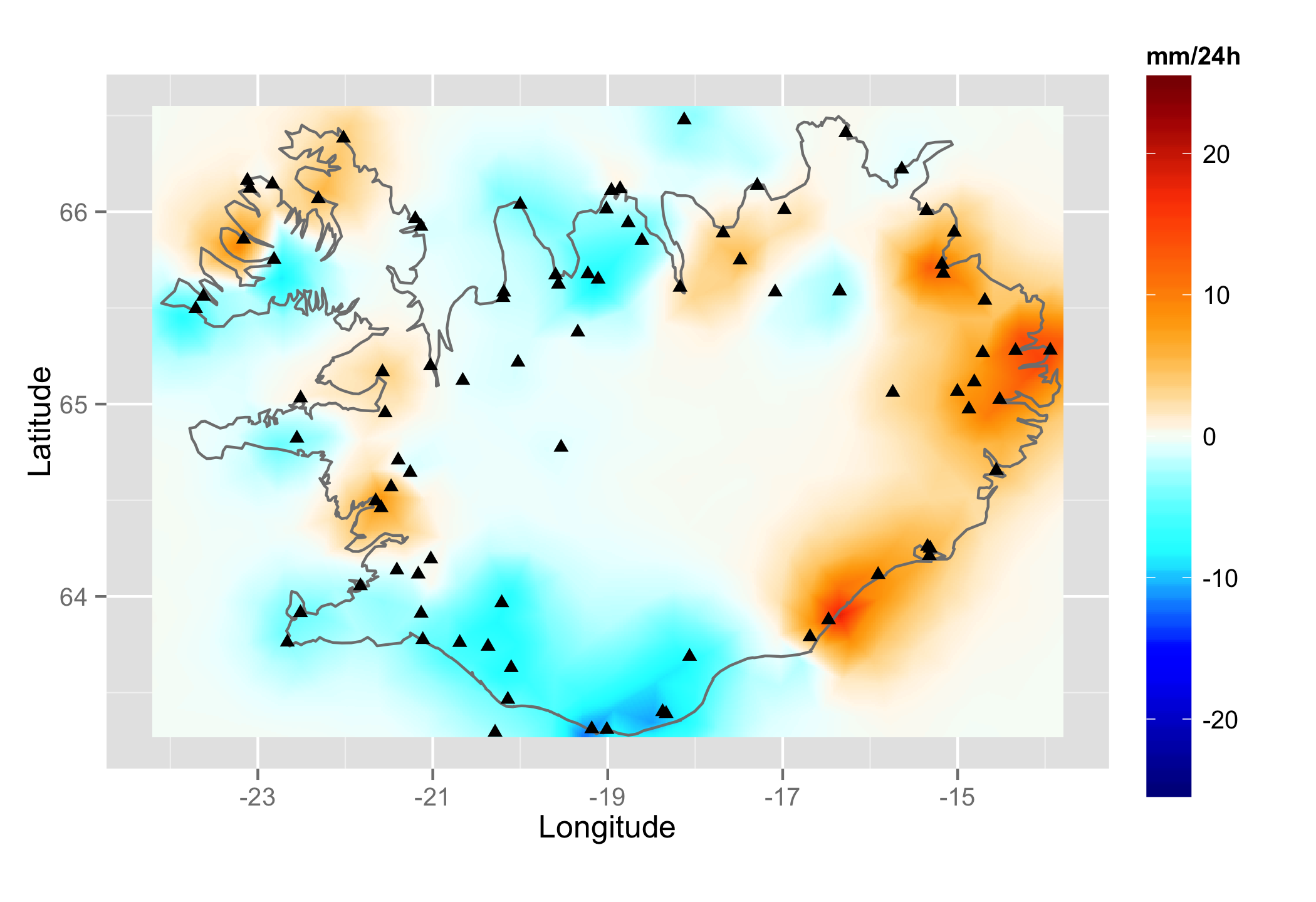}
  	\end{subfigure}
  	\begin{subfigure}[b]{0.495 \textwidth}
    	\includegraphics[width=\textwidth, trim =  15mm 25mm 30mm 20mm, clip]{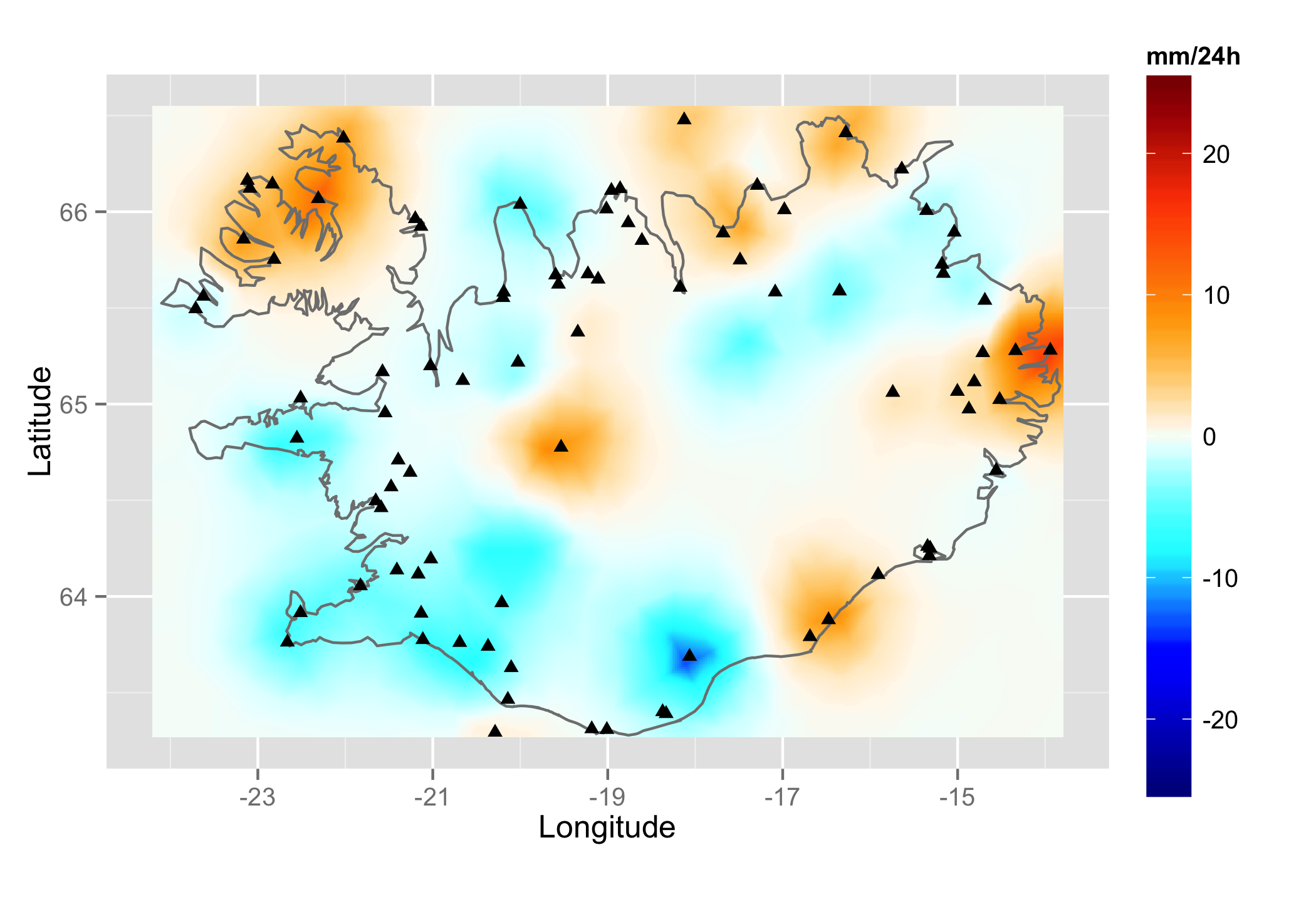}
  	\end{subfigure}

	\centering
	\begin{subfigure}[b]{0.495 \textwidth}
    	\includegraphics[width=\textwidth, trim =  15mm 25mm 30mm 20mm, clip]{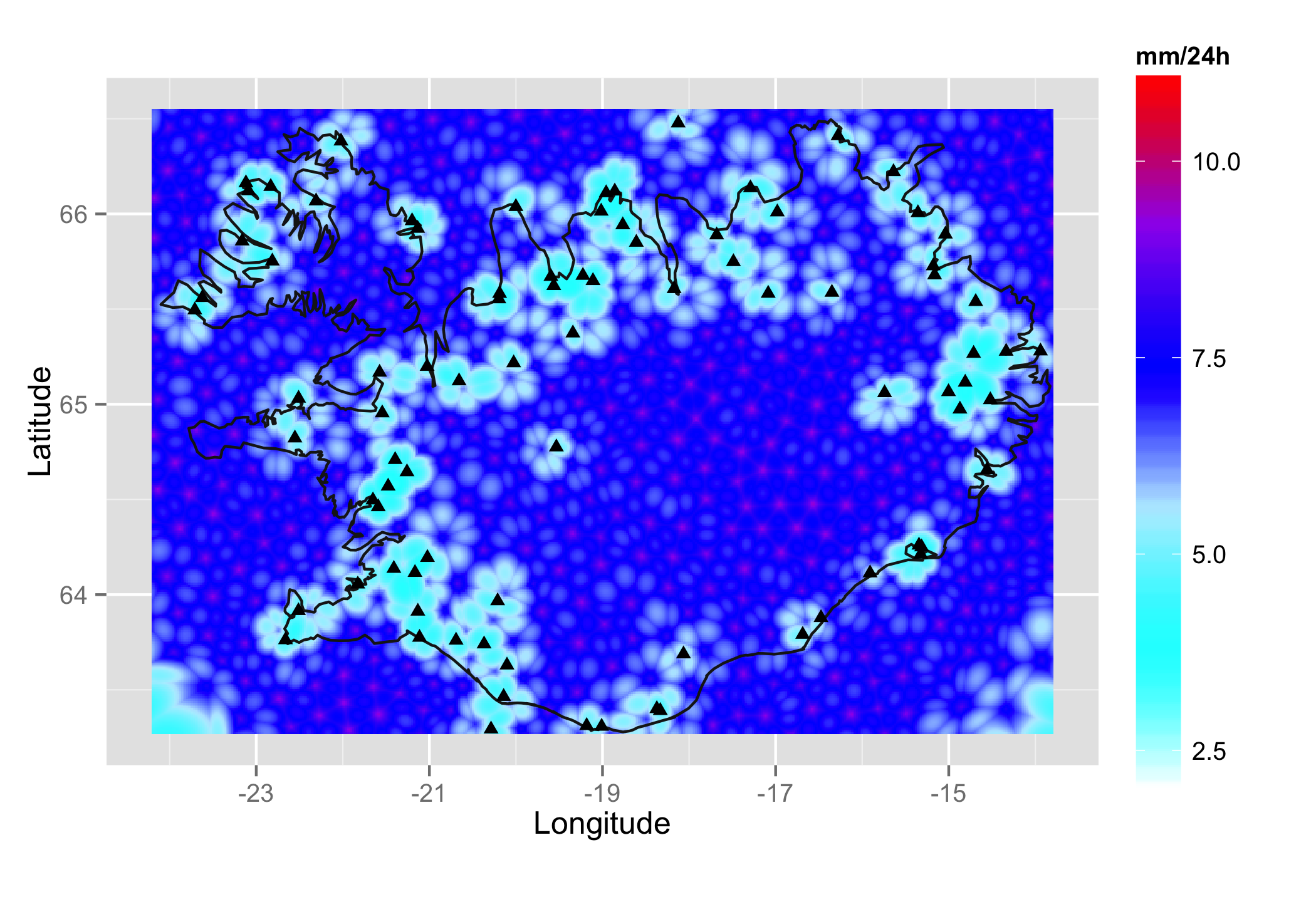}
  	\end{subfigure}
  	\begin{subfigure}[b]{0.495 \textwidth}
    	\includegraphics[width=\textwidth, trim =  15mm 25mm 30mm 20mm, clip]{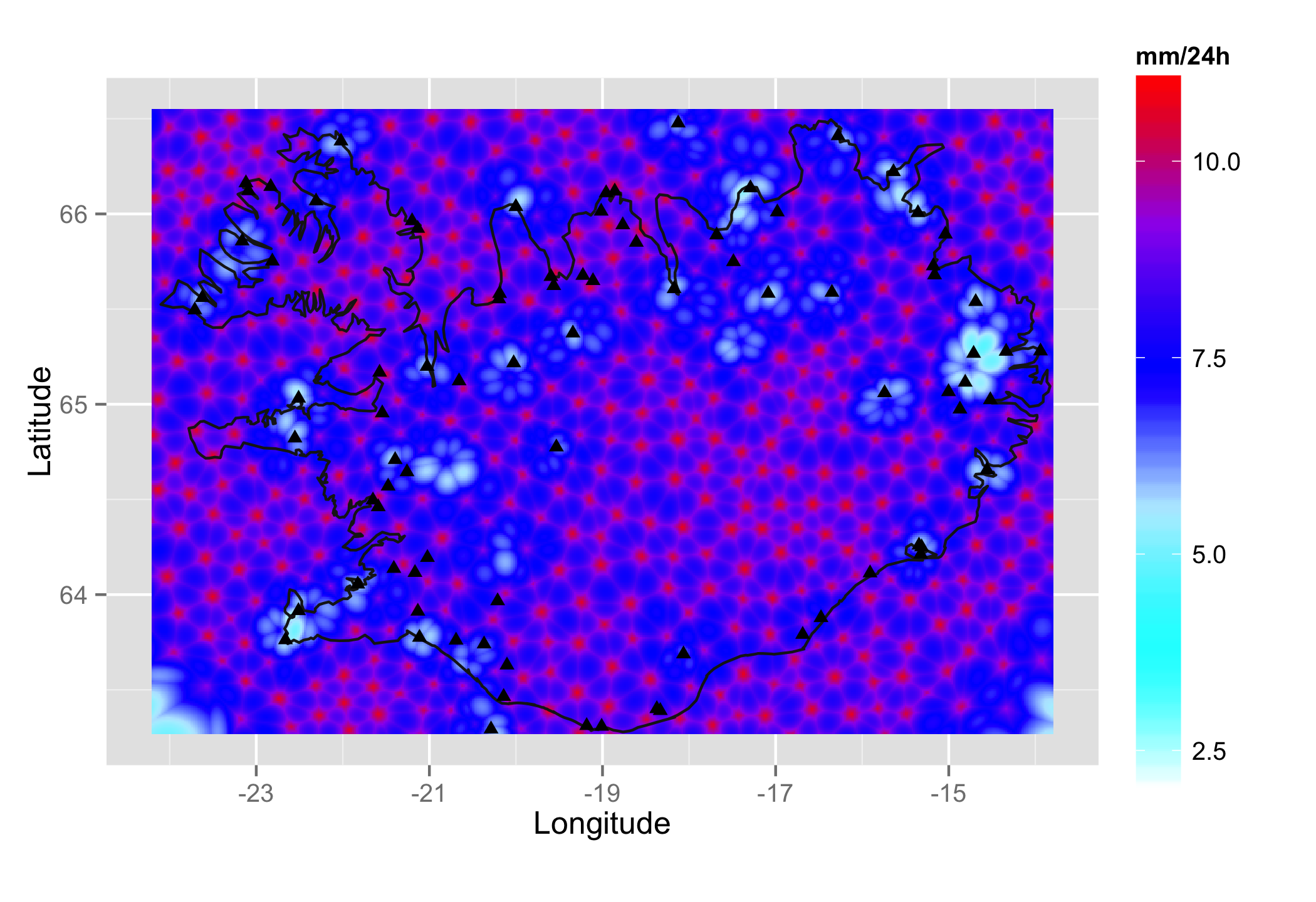}
  	\end{subfigure}
	
	\centering
	\begin{subfigure}[b]{0.495 \textwidth}
    	\includegraphics[width=\textwidth, trim =  15mm 25mm 30mm 20mm, clip]{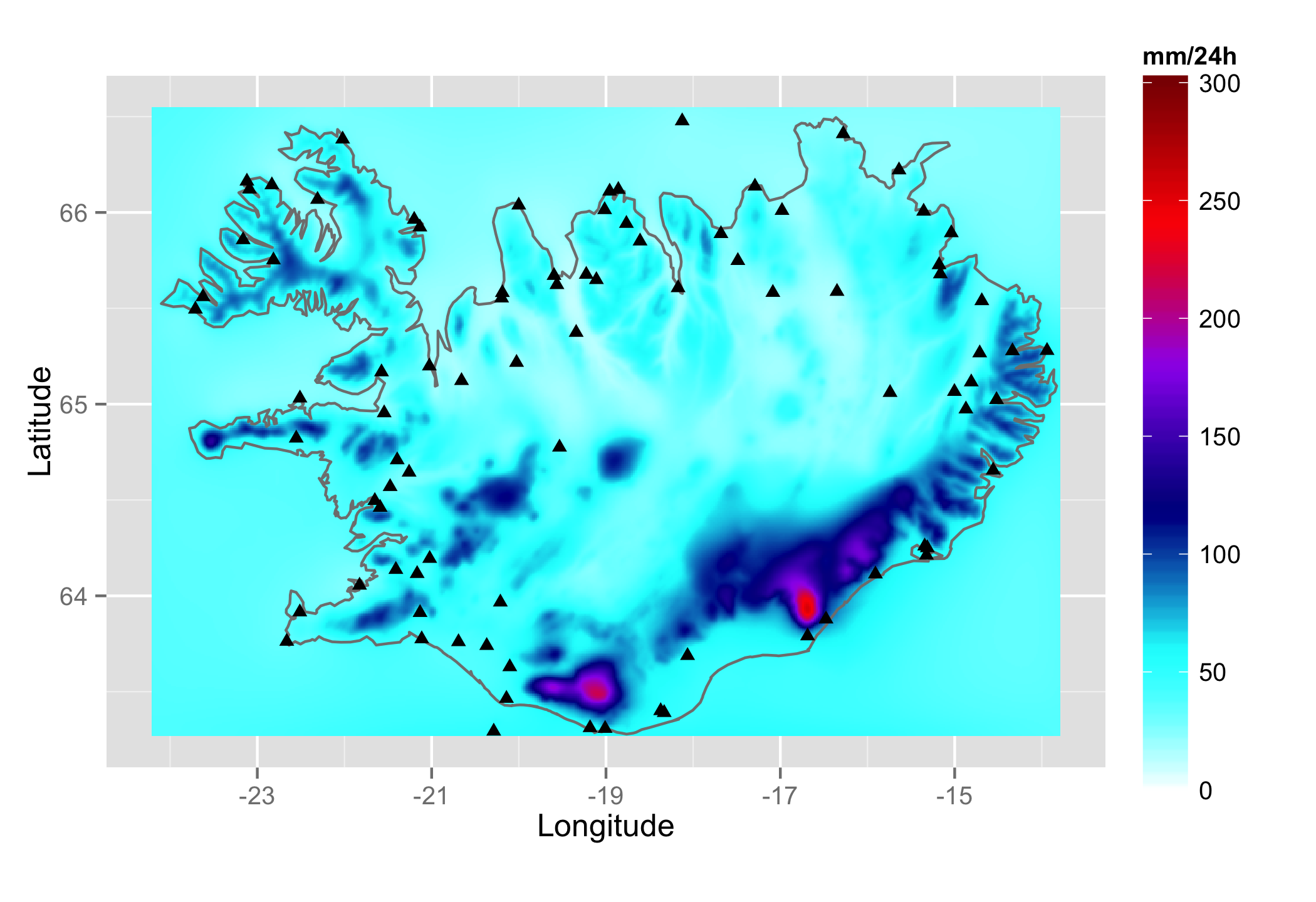}
	 \label{PostPrior_RangeEta}
  	\end{subfigure}
  	\begin{subfigure}[b]{0.495 \textwidth}
    	\includegraphics[width=\textwidth, trim =  15mm 25mm 30mm 20mm, clip]{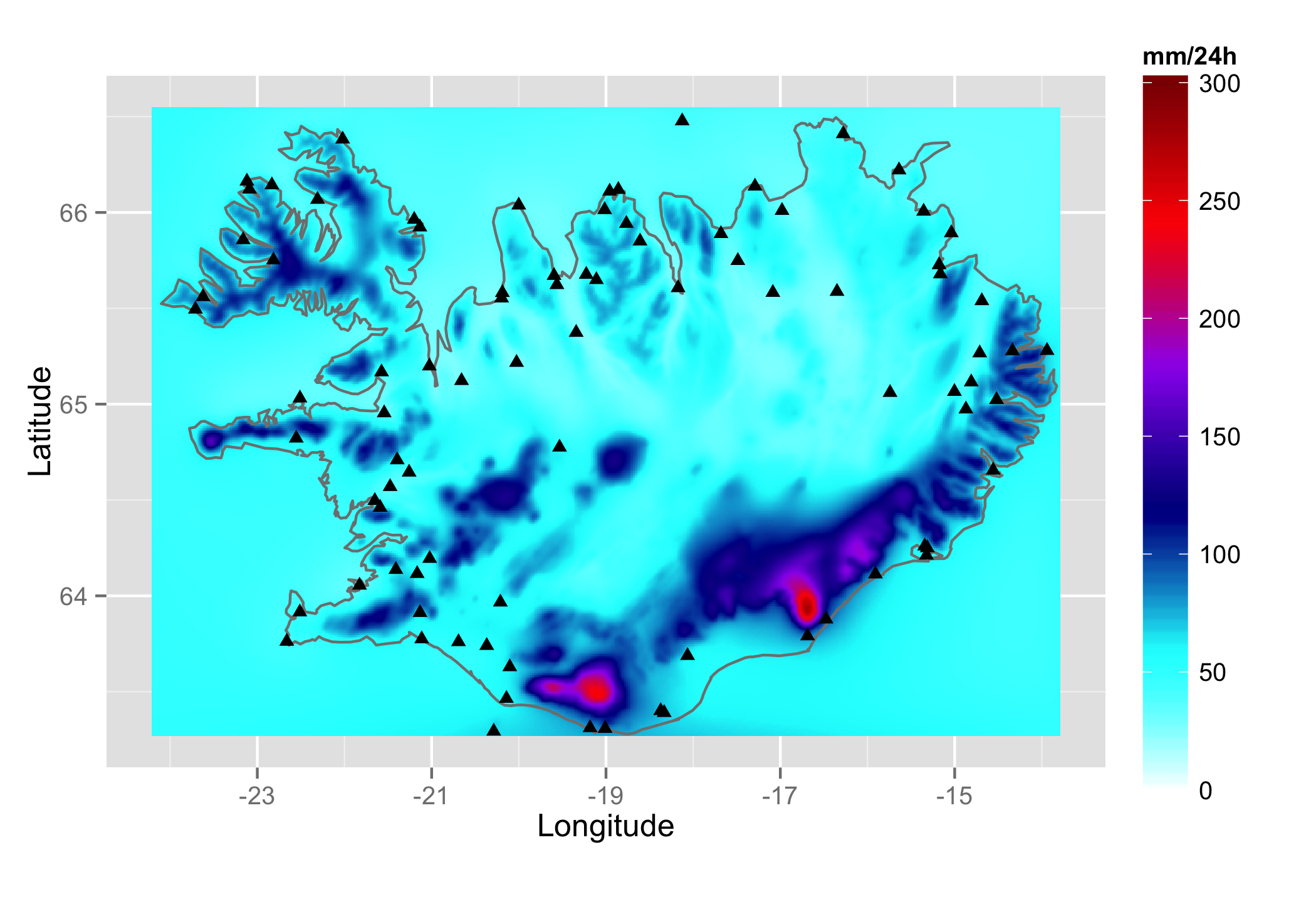}
	 \label{PostPrior_RangeTau}
  	\end{subfigure}

 \caption{The first, second and third rows show posterior mean of $\fat u_\mu$, posterior standard deviation of $\fat u_\mu$ and  posterior mean of the location parameter $\fat \mu_G$  on the regular grid $\mathcal G$, respectively. The left and right panels are based on the uncorrected and corrected data sets, respectively. }
	\label{Spatial_u_mu}
\end{figure}

The first row of Figure \ref{Spatial_u_mu} shows the spatial predictions of the posterior mean of the spatial effect $\fat u_\mu$ on the regular grid $\mathcal G$, based on (\ref{estspat}). The first row of Figure \ref{Spatial_u_mu} shows where the spatial model lowers and raises the prediction surface, meaning that the meteorological covariate overestimates and underestimates the extreme precipitation, respectively. The prediction surface is lowered the most in the south-western part of Iceland, but is raised in the south-eastern part. The spatial model yields high positive values for Kvísker, which is known to have the highest observed precipitation in Iceland; negative value in the south-eastern part close to Reykjavík and values close to zero in the interior of Iceland where there are no observational sites. 

The second row of Figure \ref{Spatial_u_mu} shows the spatial prediction for the standard deviation of the spatial effect $\fat u_\mu$ on the regular grid $\mathcal G$. As expected, the standard deviation is less where there is more data forming sinks in the standard deviation near the observation points. The standard deviation increases at points further away from the observational sites. An interesting artifact of the SPDE modeling also appears. The estimates for the standard deviation are lower for points inside the triangles than in their corresponding edges or vertices in the areas where there are no data. This happens due to the linear basis function and because the standard deviation is estimated based on (\ref{estspat}). 

The third row of Figure \ref{Spatial_u_mu} shows the spatial prediction for the posterior mean of the location parameter $\fat \mu_{\mathcal G}$, based on (\ref{muspat}), on the regular grid $\mathcal G$. The figure shows that the mean of the extreme precipitation is at its highest in the south-eastern parts of Iceland, in the vicinity of the southern side of Vatnajökull glacier. This is to be expected, as the spatial gradient at the southern side of Vatnajökull increases rapidly moving from the nearby coastline to the top of the glacier. Due to these topographical properties and the fact that humid air blows in from the southern shoreline towards the roots of the glacier, the physical law of orographic precipitation predicts high precipitation. However, the areas north of Vatnajökull and in the middle of the country are known to be in a rain shadow by the same meteorological law. That is again in line with the results seen in Figure \ref{Spatial_u_mu}.

\begin{figure}[htp]
	\centering
	\begin{subfigure}[b]{0.495 \textwidth}
    	\includegraphics[width=\textwidth, trim =  15mm 25mm 30mm 20mm, clip]{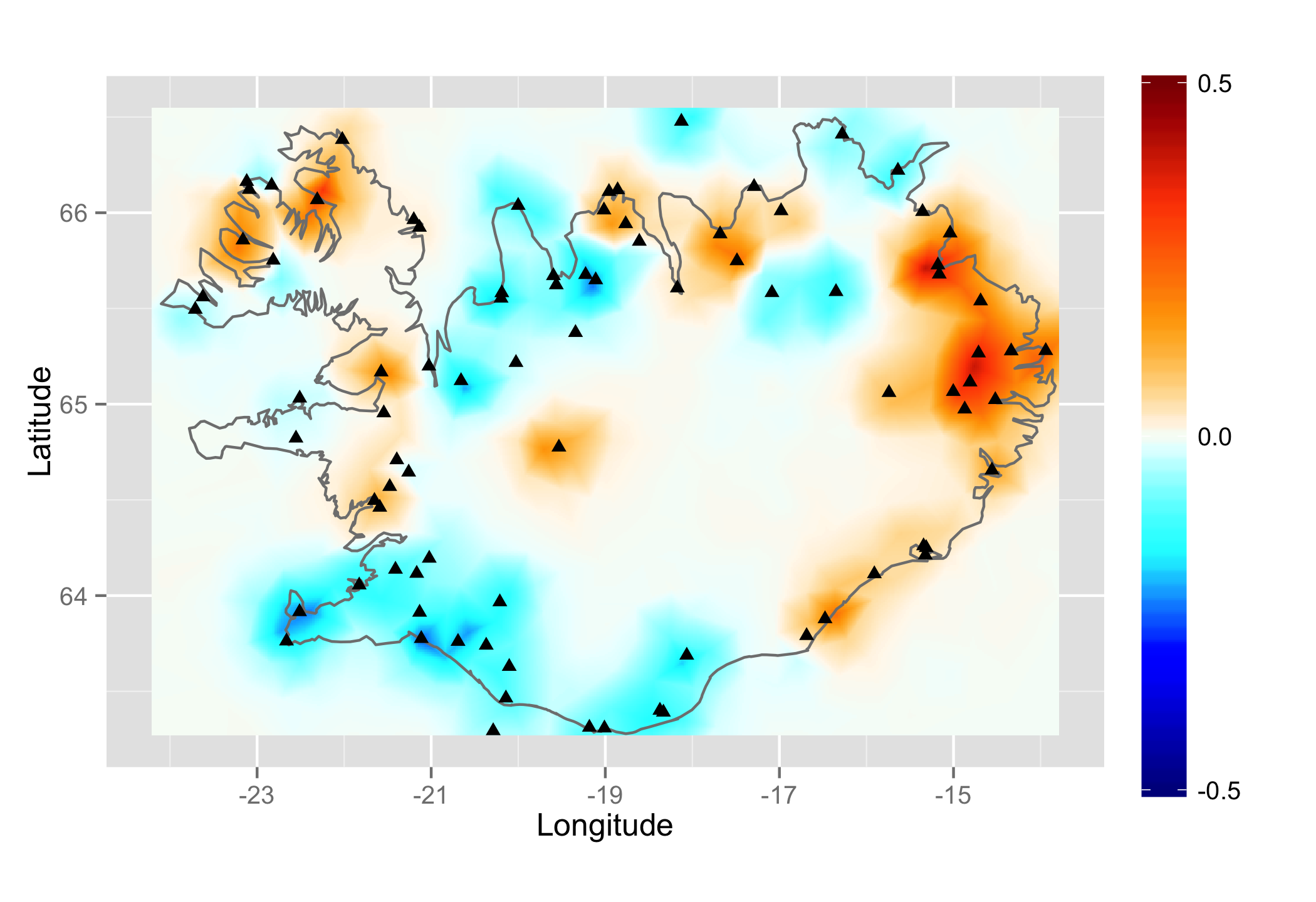}
  	\end{subfigure}
  	\begin{subfigure}[b]{0.495 \textwidth}
    	\includegraphics[width=\textwidth, trim =  15mm 25mm 30mm 20mm, clip]{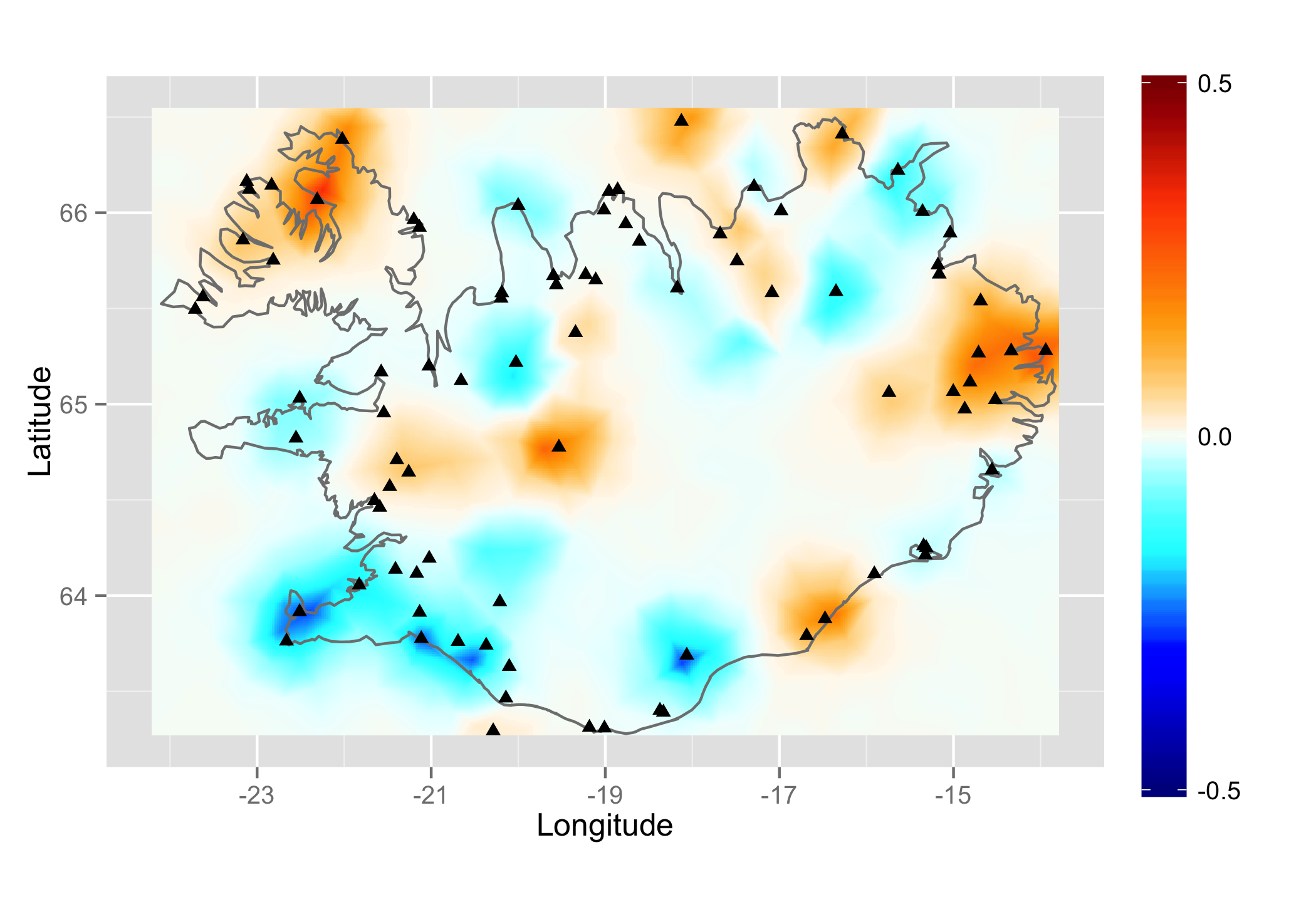}
  	\end{subfigure}

	\centering
	\begin{subfigure}[b]{0.495 \textwidth}
    	\includegraphics[width=\textwidth, trim =  15mm 25mm 30mm 20mm, clip]{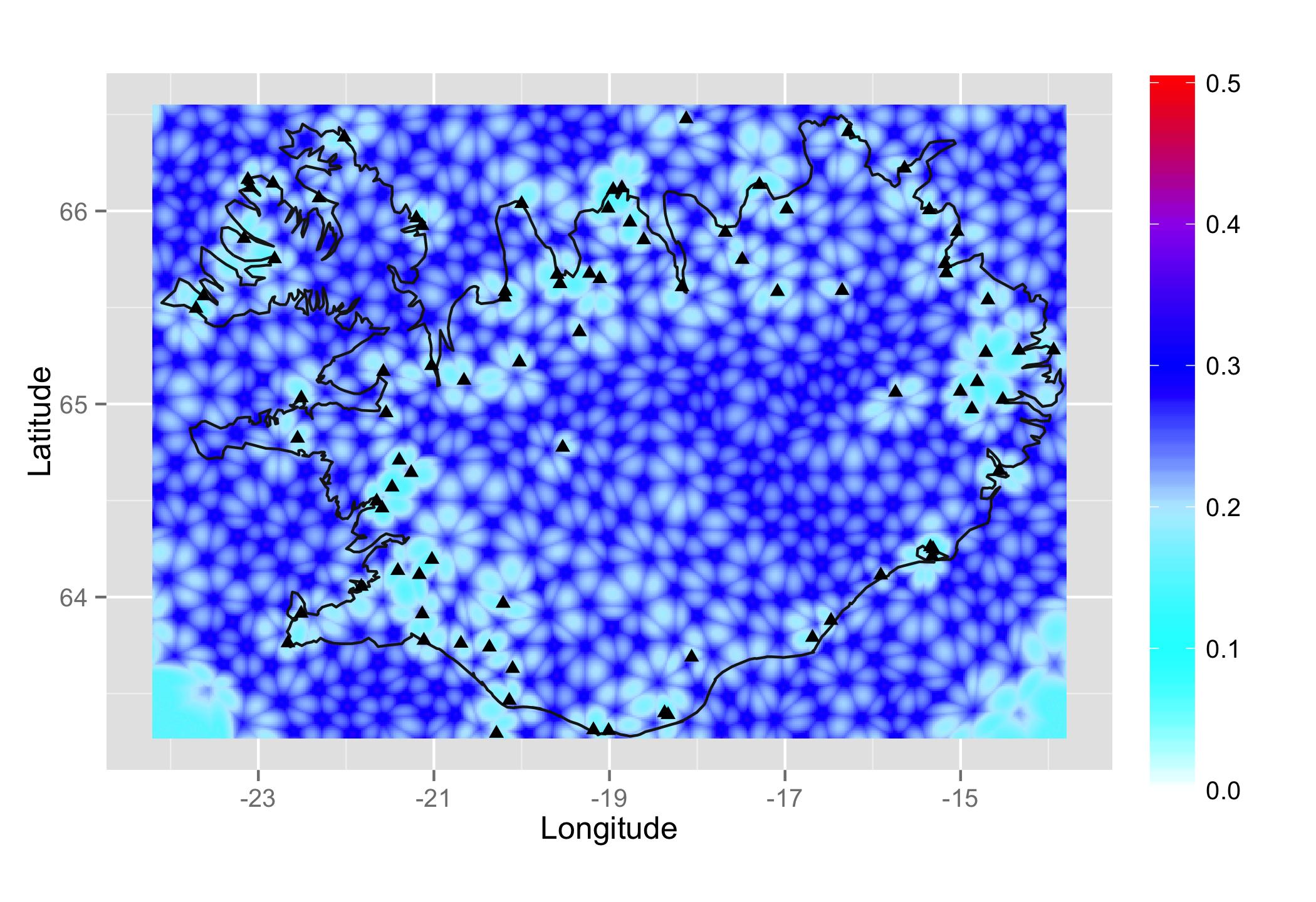}
  	\end{subfigure}
  	\begin{subfigure}[b]{0.495 \textwidth}
    	\includegraphics[width=\textwidth, trim =  15mm 25mm 30mm 20mm, clip]{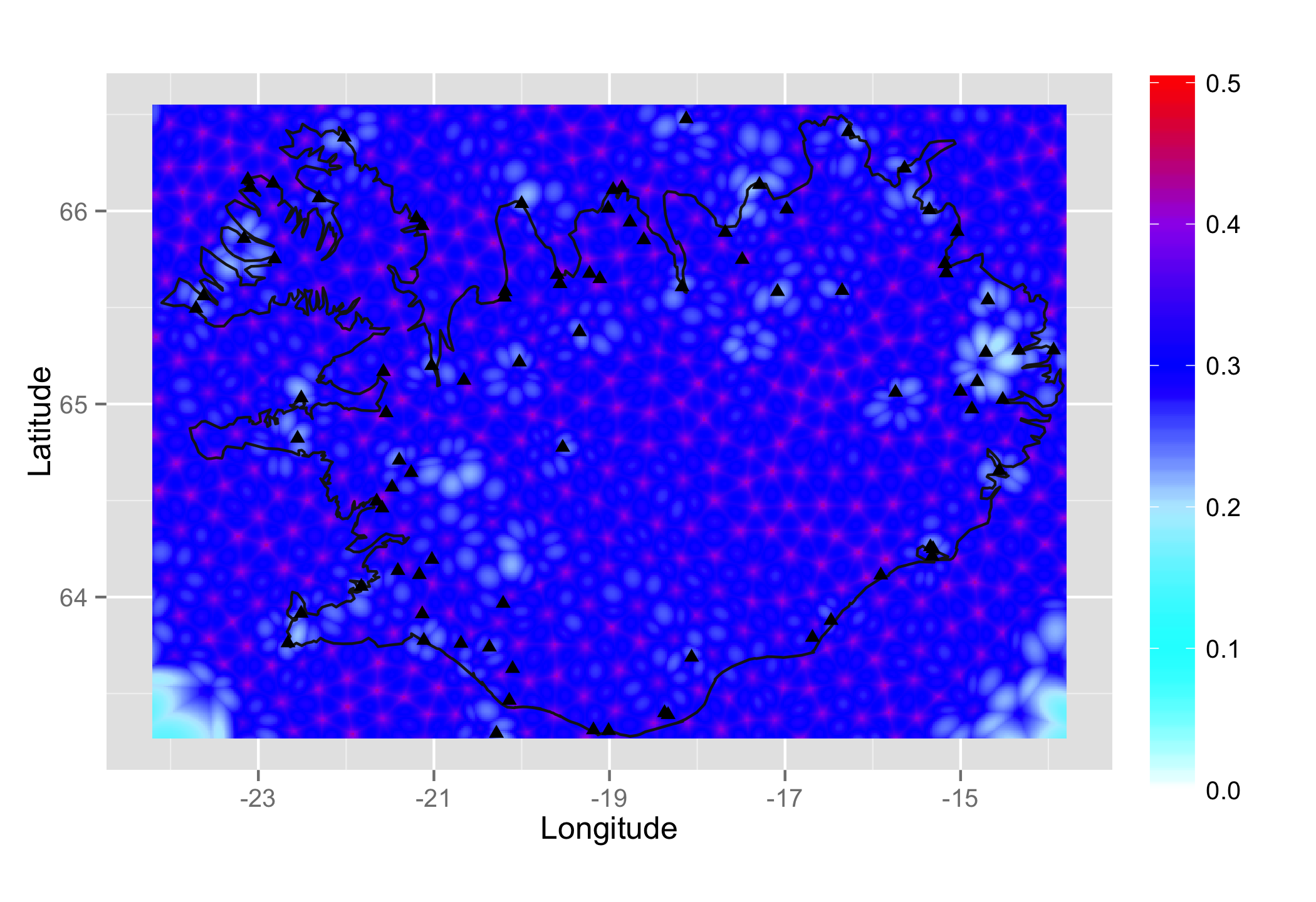}
  	\end{subfigure}
	
	\centering
	\begin{subfigure}[b]{0.495 \textwidth}
    	\includegraphics[width=\textwidth, trim =  15mm 25mm 30mm 20mm, clip]{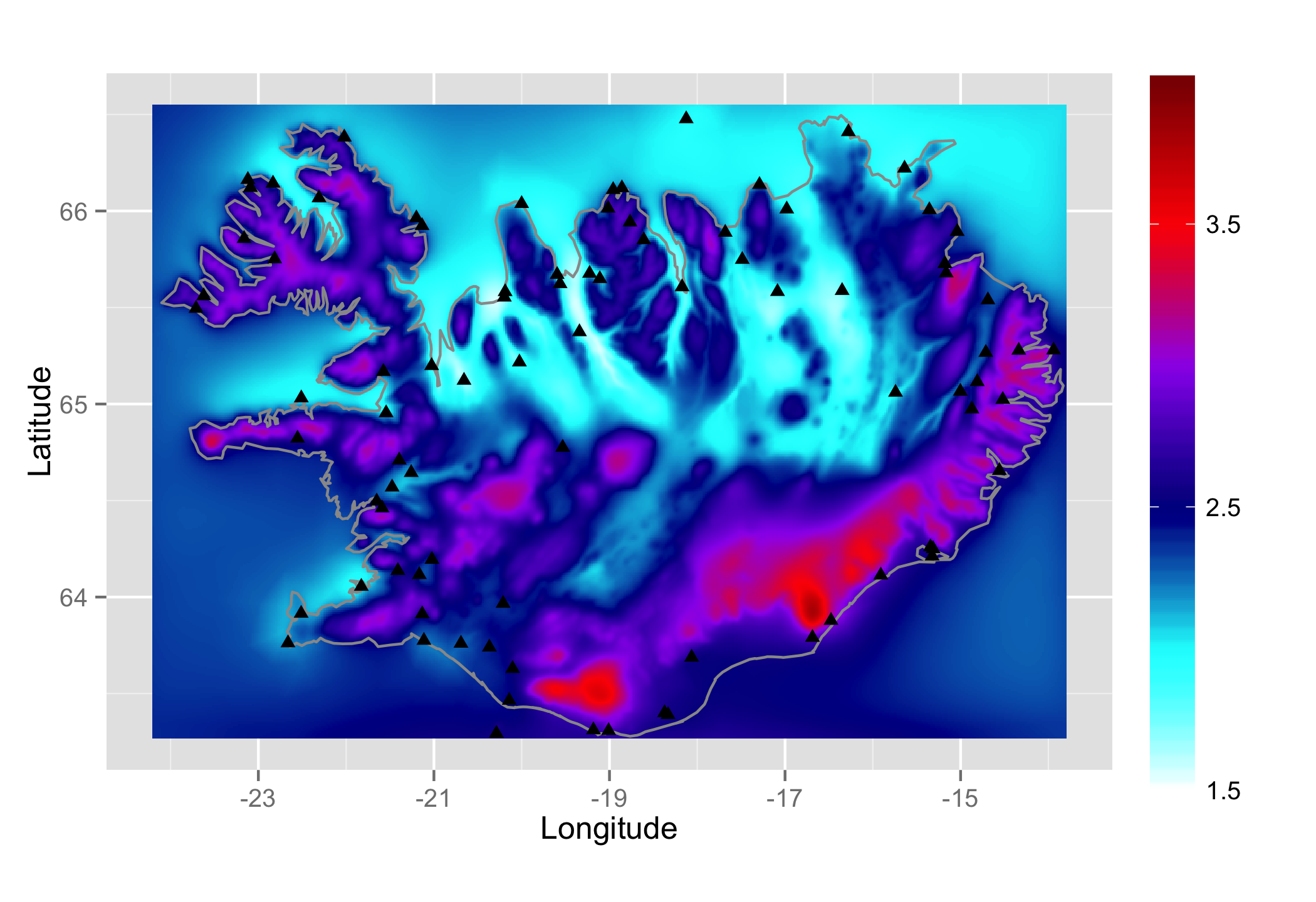}
	 \label{PostPrior_RangeEta}
  	\end{subfigure}
  	\begin{subfigure}[b]{0.495 \textwidth}
    	\includegraphics[width=\textwidth, trim =  15mm 25mm 30mm 20mm, clip]{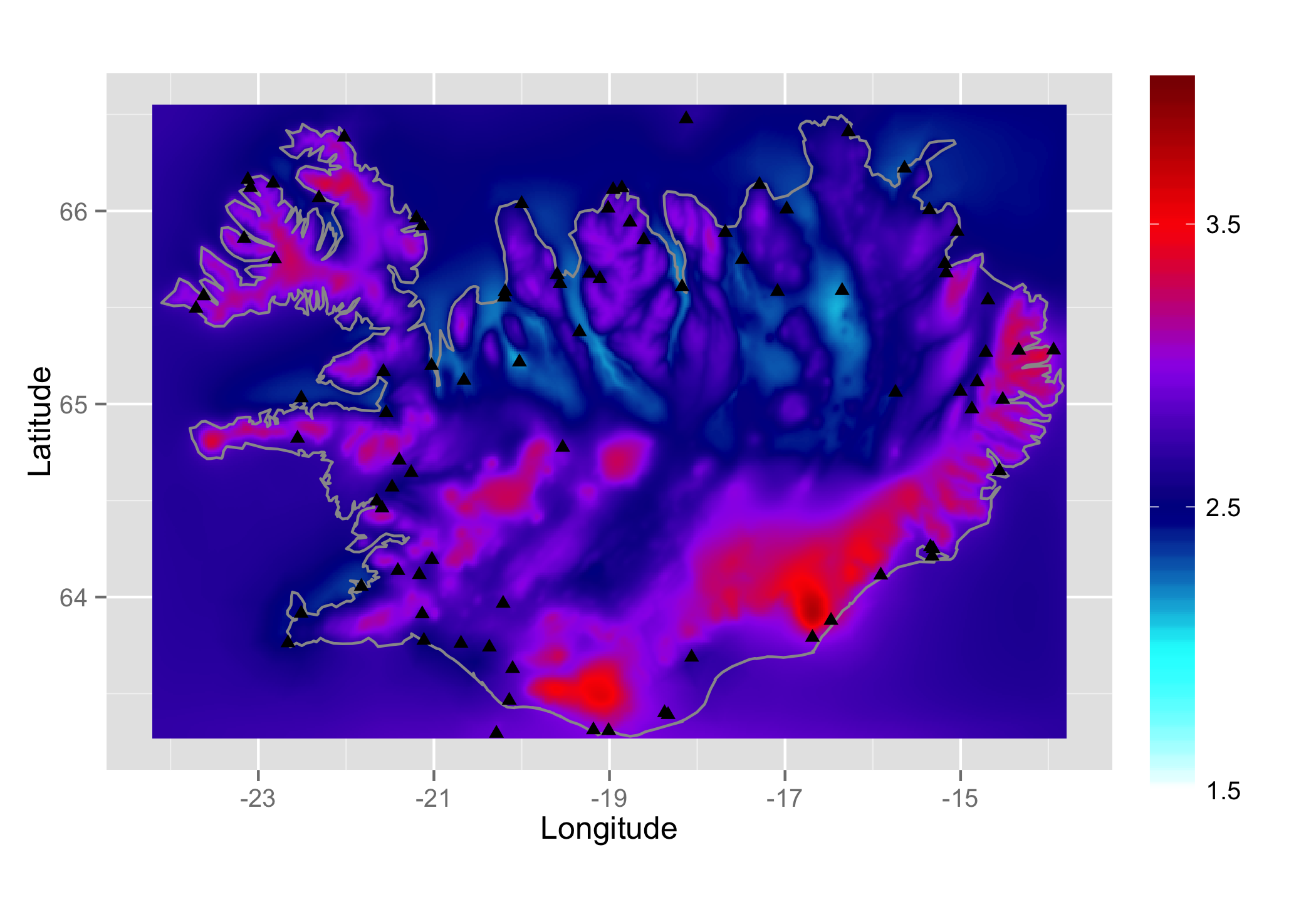}
	 \label{PostPrior_RangeTau}
  	\end{subfigure}

 \caption{The first, second and third rows show posterior mean of $\fat u_\tau$, posterior standard deviation of $\fat u_\tau$ and  posterior mean of the location parameter $\fat \tau_G$  on the regular grid $\mathcal G$, respectively. The left and right panels are based on the uncorrected and corrected data sets, respectively. }
	\label{Spatial_u_tau}
\end{figure}

Figure \ref{Spatial_u_tau} shows the spatial prediction for the scale parameter on a logarithmic scale, and is arranged in the same manner as Figure \ref{Spatial_u_mu}. In the first row, we see that the spatial model raises the spatial surface in the south-eastern part and lowers it in the south-western part. In the second row, similar results appear for the standard deviation as for the location parameter. In the third row, we can see that the estimates for the scale parameter are highest along the south-eastern coastline. 

\begin{figure}[htp]
	\centering
	\begin{subfigure}[b]{0.495 \textwidth}
    	\includegraphics[width=\textwidth, trim =  15mm 25mm 30mm 20mm, clip]{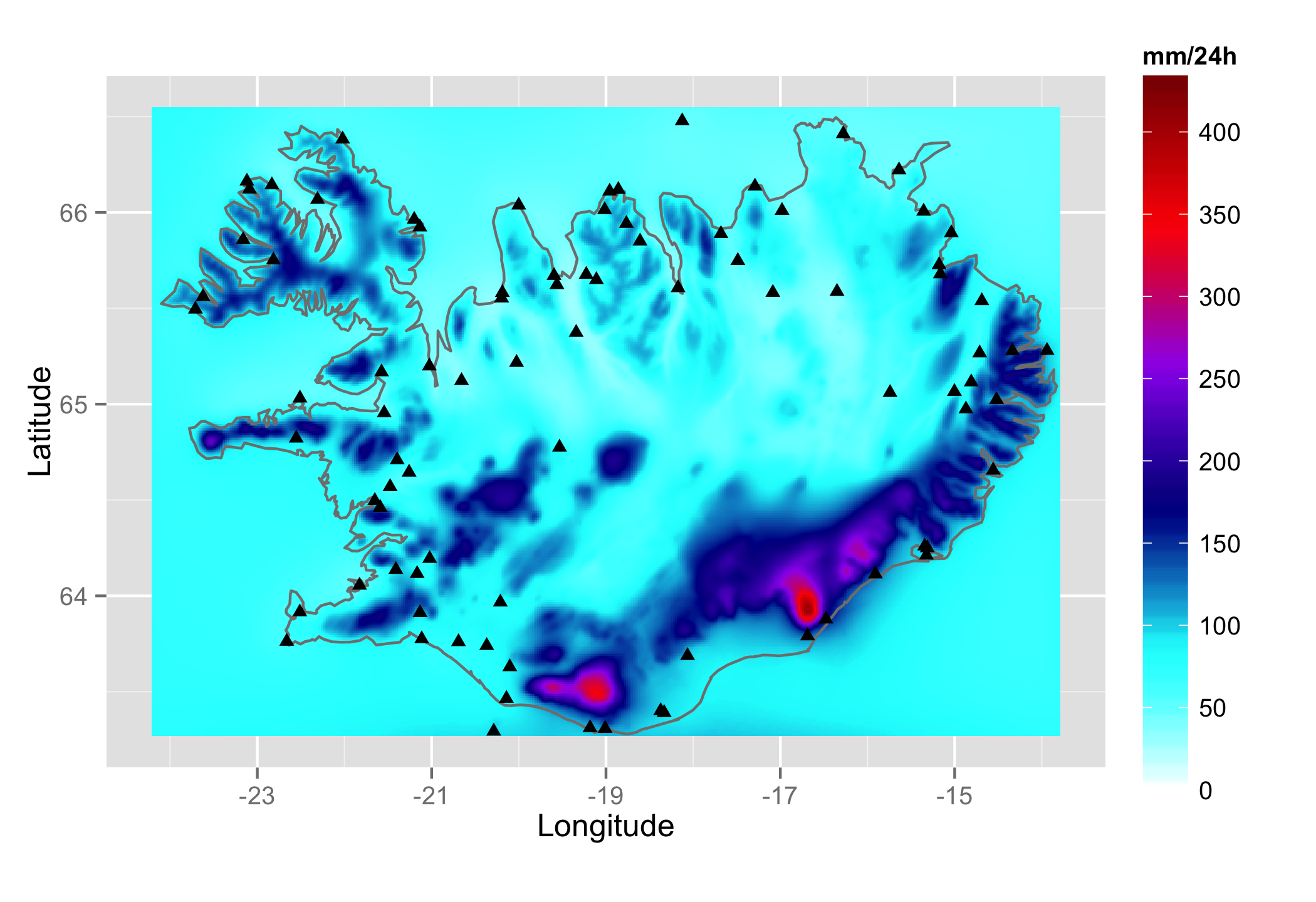}
  	\end{subfigure}
  	\begin{subfigure}[b]{0.495 \textwidth}
    	\includegraphics[width=\textwidth, trim =  15mm 25mm 30mm 20mm, clip]{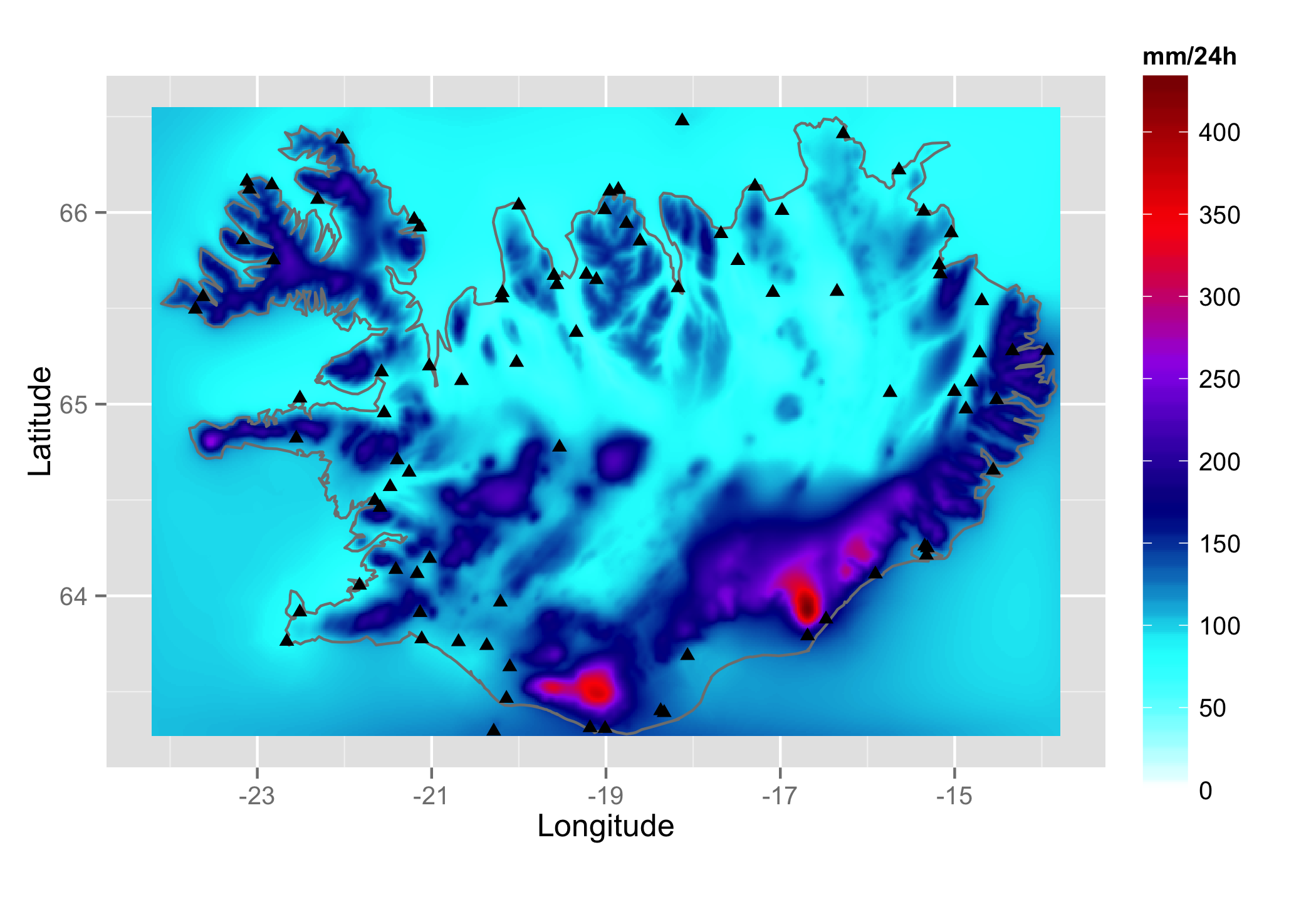}
  	\end{subfigure}

 \caption{The posterior mean of the 0.95 quantile of the generalized extreme value distribution. The left and right panels are based on the uncorrected and corrected data sets, respectively. }
	\label{Kvant95}
\end{figure}

Spatial predictions for the posterior mean of the 0.95 quantile of the generalized extreme value distribution can be seen in Figure \ref{Kvant95}. The predictions are based on the methods discussed in Section \ref{quantpre} and can be interpreted as the 20-year precipitation event. The results reflect the previous posterior results about the location and scale parameters. For example, the highest predicted 20-year precipitation events are along the south and south-eastern coastlines, while the lowest predicted events are in the interior of Iceland. The maximum of the predicted values is at the southern side of Hvannadalshnjúkur and is approximately 400 mm per 24 hours, based on the uncorrected data set. The peak of Hvannadalshnjúkur, which is a part Vatnajökull, is the highest point of Iceland and is approximately 10 km north west of the observational site Kvísker. These results demonstrate the effects of orographic precipitation as Kvísker is around 30 m above sea level and the highest point of Hvannadalshnjúkur is roughly 2110 m above sea level.  Furthermore, the 20-year precipitation event in Reykjavík and surroundings is predicted close to 54 mm per 24 hours and in the vicinity of Akureyri it is around 42 mm, based on the uncorrected data set.

\section{Discussion}

The statistical methodology presented in this paper is in line with 'going beyond mean regression' discussed by \cite{kneib2013beyond}. Working within the LGM framework yields a flexible and computationally efficient way of doing statistical modeling, rather than only doing mean regression. Furthermore, the framework also allows scientific knowledge to be assimilated into a statistical model in a natural way, as presented above. 

The presented modeling framework is modular by design. This means that other likelihoods can be chosen for the observations without the need to change the structure of the spatial model at the latent level. A different likelihood choice for extreme precipitation is, for example, the peek over threshold methods with the Generalized Pareto Distribution as presented in \citep{cooley2007bayesian}. The advantages of that approach is that more of the data can be incorporated into the modeling, depending on the choice of threshold. However, this approach was not chosen for the modeling in this paper, as choosing a common threshold for each station is somewhat unrealistic for the data sets explored in this as this paper. For example, in Figure \ref{FIG:timseS}, it can be seen that highest observed value in Reykjavík is lower than the lowest observed value in Kvísker. 

Another likelihoods choices are copula likelihoods with marginals based on the generalized extreme value distribution. Likelihoods constructed with the $t$-copula have been explored for example by \cite{davison2012statistical} to model extreme precipitation and present an appealing choice to model the probabilistic dependence of the observations. The approach is well suited for simulating spatial realizations of the extreme precipitation for the next year or unobserved sites for an observed year. Although the copula likelihood can be implemented within the presented modeling framework, there were two main reasons why it was not chosen in this paper. First, 
the main purpose is to give spatial predictions of marginal quantiles using the SPDE approach but not to simulate spatial realizations of the extremal surfaces. Secondly, many of the observational sites had missing observations, which presents further computational difficulties with the copula based likelihood.

At the latent level of the model, we have implemented a stationary SPDE spatial models. However, the results indicated a non-stationary behavior in some of the mountainous regions, for example near Kvísker, as discussed in Section \ref{PE}. Non-stationary SPDE spatial models have been proposed \citep{fuglstad2013non, Ingebrigtsen2013}, and yield an appealing modeling option for non-stationary spatial fields. However, due to the sparsity of the observational sites, these models were not implemented.

\section*{Acknowledgements}

The authors would like to thank the University of Iceland Doctoral Fund, University of Iceland Research Fund and Landsvirkjun Energy Research Fund which supported the research. The authors would also like to thank the Icelandic Meteorological Office, in particular, Dr. Philippe Crochet for providing the data and valuable discussions. The authors give their thanks to the Nordic Network on Statistical Approaches to Regional Climate Models for Adaptation (SARMA), especially Prof. Peter Guttorp, for providing travel support. Furthermore, the authors give their thanks to the  Department of Mathematical Sciences at the Norwegian University of Science and Technology for hosting Óli Páll Geirsson several times, and special gratitude to Prof. Håvard Rue for his invitation and valuable conversations.

\bibliographystyle{plain}
\bibliography{bibfile}

\newpage
\section{Appendix}

\subsection{Convergence diagnostic plots}

\end{document}